\documentclass[12pt,preprint]{aastex}

\shorttitle{H.E.S.S. observations of the supernova remnant RX J0852.0-4622}
\shortauthors{F.~A. Aharonian et al.}

\begin{document}

\sloppy

%% LaTeX will automatically break titles if they run longer than
%% one line. However, you may use \\ to force a line break if
%% you desire.

\title{H.E.S.S. observations of the supernova remnant RX~J0852.0-4622: 
shell-type morphology and spectrum of a widely extended VHE gamma-ray 
source.}

%% Use \author, \affil, and the \and command to format
%% author and affiliation information.
%% Note that \email has replaced the old \authoremail command
%% from AASTeX v4.0. You can use \email to mark an email address
%% anywhere in the paper, not just in the front matter.
%% As in the title, use \\ to force line breaks.

\author{F. Aharonian~\altaffilmark{1},
 A.G.~Akhperjanian~\altaffilmark{2},
 A.R.~Bazer-Bachi~\altaffilmark{3},
 M.~Beilicke~\altaffilmark{4},
 W.~Benbow~\altaffilmark{1},
 D.~Berge~\altaffilmark{1,a},
 K.~Bernl\"ohr~\altaffilmark{1,5},
 C.~Boisson~\altaffilmark{6},
 O.~Bolz~\altaffilmark{1},
 V.~Borrel~\altaffilmark{3}
% ...
}

% ...
\altaffiltext{1}{
Max-Planck-Institut f\"ur Kernphysik, Heidelberg, Germany}
\altaffiltext{2}{Yerevan Physics Institute, Yerevan, Armenia}
\altaffiltext{3}{Centre d'Etude Spatiale des Rayonnements, CNRS/UPS,
Toulouse, France} \altaffiltext{4}{Universit\"at Hamburg, Institut
f\"ur Experimentalphysik, Hamburg, Germany} \altaffiltext{5}{Institut
f\"ur Physik, Humboldt-Universit\"at zu Berlin, Germany}
\altaffiltext{6}{LUTH, UMR 8102 du CNRS, Observatoire de Paris,
Section de Meudon, France}
% ...
\altaffiltext{a}{now at CERN, Geneva, Switzerland}
% ...
%% Notice that each of these authors has alternate affiliations, which
%% are identified by the \altaffilmark after each name.  Specify alternate
%% affiliation information with \altaffiltext, with one command per each
%% affiliation.

%% Mark off your abstract in the ``abstract'' environment. In the manuscript
%% style, abstract will output a Received/Accepted line after the
%% title and affiliation information. No date will appear since the author
%% does not have this information. The dates will be filled in by the
%% editorial office after submission.

\begin{abstract}
The shell-type supernova remnant RX~J0852.0-4622 was observed with the High Energy Stereoscopic System (H.E.S.S.) of Atmospheric Cherenkov Telescopes between December 2004 and May 2005 for a total observation time of 33 hours, above an average gamma-ray energy threshold of 250~GeV. The angular resolution of $\sim 0.06^{\circ}$ (for events triggering 3 or 4 telescopes) and the large field of view of H.E.S.S. ($5^{\circ}$ diameter) are well adapted to studying the morphology of the object in very high energy gamma-rays, which exhibits a remarkably thin shell very similar to the features observed in the radio range and in X-rays. The spectral analysis of the source from 300~GeV to 20~TeV is also presented. Finally, the possible origins of the very high energy gamma-ray emission (Inverse Compton scattering by electrons or the decay of neutral pions produced by proton interactions) are discussed, on the basis of morphological and spectral features obtained at different wavelengths.
\end{abstract}

%% Keywords should appear after the \end{abstract} command. The uncommented
%% example has been keyed in ApJ style. See the instructions to authors
%% for the journal to which you are submitting your paper to determine
%% what keyword punctuation is appropriate.


%% From the front matter, we move on to the body of the paper.
%% In the first two sections, notice the use of the natbib \citep
%% and \citet commands to identify citations.  The citations are
%% tied to the reference list via symbolic KEYs. The KEY corresponds
%% to the KEY in the \bibitem in the reference list below. We have
%% chosen the first three characters of the first author's name plus
%% the last two numeral of the year of publication as our KEY for
%% each reference.

%% Authors who wish to have the most important objects in their paper
%% linked in the electronic edition to a data center may do so by tagging
%% their objects with \objectname{} or \object{}.  Each macro takes the
%% object name as its required argument. The optional, square-bracket 
%% argument should be used in cases where the data center identification
%% differs from what is to be printed in the paper.  The text appearing 
%% in curly braces is what will appear in print in the published paper. 
%% If the object name is recognized by the data centers, it will be linked
%% in the electronic edition to the object data available at the data centers  

\section{Introduction}
Shell-type supernova remnants (SNR) have long been considered as
prime candidates for accelerating cosmic rays up to at least 
100~TeV, but until recently, this statement was only supported by
indirect evidence, namely non-thermal X-ray emission interpreted as
synchrotron radiation from very-high-energy (VHE) electrons in a few
objects~\citep{koyama1,koyama2}. A more direct proof is
provided by the emission of high-energy gamma-rays produced either
by Inverse Compton (IC) scattering of high-energy electrons on
ambient photons or by nuclear interactions of high-energy protons or
ions in the interstellar medium and subsequent $\pi^0$ meson decays.
However, in the $100 \, \rm{MeV} - 30 \, \rm{GeV}$ energy range, the Energetic Gamma-Ray
Experiment Telescope (EGRET) onboard the Compton Gamma-Ray
Observatory could not provide an unambiguous detection of a
shell-type SNR, due to its poor angular resolution and to the
difficulty of separating signals of extended objects from the
diffuse galactic gamma-ray background. In the very-high-energy range
($>$ 200 GeV) on the other hand, the situation is more favorable
\citep{drury}: recent Imaging Atmospheric Cherenkov Telescopes have
achieved angular resolutions of the order of a few arc minutes and
the diffuse background is expected to decrease more rapidly with
energy than the gamma-ray signal. The first confirmed gamma-ray
signal from a shell-type SNR was that of RX~J1713.7-3946 detected by
the CANGAROO-I and CANGAROO-II experiments~\citep{muraishi,enomoto} 
as well as by the H.E.S.S. collaboration~\citep{hessrxjnature}. 
The latter experiment provided the first
detailed morphological and spectral study of this source~\citep{HESSRXJ}. 
A second shell-type supernova remnant,
RX~J0852.0-4622 (also named G266.2-1.2), was recently detected by
Cherenkov telescopes: the announcement of a signal from the
north-western part by the CANGAROO collaboration~\citep{katagiri} was
shortly followed by the publication of a complete gamma-ray map of
this object by the H.E.S.S. collaboration~\citep{HESSVelaJr}, 
 making it the largest extended 
source (2$^{\circ}$ angular diameter) ever resolved by a 
Cherenkov telescope. This previous 
H.E.S.S. result was obtained from a short observation campaign (3.2
hours) in 2004. In this article, we present the results of much
longer observations of this source in 2005 ($\sim$ 20 hours) with
the full H.E.S.S. array.\\

%RX~J0852.0-4622 presents some remarkable features. First, it is the
%largest extended source (2$^{\circ}$ angular diameter) ever detected by a
%Cherenkov telescope; the large field of view of the H.E.S.S.
%telescopes (5$^{\circ}$ diameter) is therefore well adapted to its
%study. Furthermore, the X-ray emission of
%RX~J0852.0-4622~\citep{aschen98}, like that of RX~J1713-3946, is
%dominated by a non-thermal component and this object is also a very
%weak radio-emitter~\citep{combi,duncan}. On the other hand, it
RX~J0852.0-4622 is located in the south-eastern corner of the Vela SNR and its
study in X-rays as well as in radio is complicated by the
superposition of the highly structured emission of the Vela remnant. Its
discovery in the ROSAT all-sky survey~\citep{aschen98} relied on the
restriction to energies greater than 1.3~keV where the signal stands
out above the soft thermal emission from the Vela SNR. In X-rays,
RX~J0852.0-4622 appears as a roughly circular disk with a diameter
of 2$^{\circ}$ with a brightening towards the north-western, western
and south-eastern rims. 
%These regions have been observed in more
%detail by the ASCA~\citep{tsunemi,slane}, XMM-Newton~\citep{iyudin}, 
%and Chandra~\citep{bamba} satellites and their
%corresponding X-ray spectra have been confirmed to be mainly
%non-thermal. \\
Since its discovery, its distance and age have been a
matter of controversy. Low values of these quantities have been
inferred from the detection by COMPTEL~\citep{iyudin0} of the
1.157~MeV gamma-ray line of $^{44}$Ca due to the decay chain
$^{44}$Ti$\, \rightarrow \, ^{44}$Sc$\, \rightarrow \, ^{44}$Ca
characterized by the $^{44}$Ti lifetime, whose average value, based on
measurements by independent groups, is $86.6 \pm 1.4$~years \citep{hashimoto}.
%of about 90 years; 
On the basis of the $^{44}$Ti yield and of the X-ray diameter, an age of
$\sim$ 680 yr and a distance of $\sim$ 200~pc, thus close to that of
the Vela remnant, was proposed~\citep{aschen99}. It should be noted that this result was 
obtained by adopting a mean expansion velocity of 5000~km~s$^{-1}$ based on the
assumption 
of a purely thermal X-ray spectrum. However, further observations of the
brightest parts of the shell by ASCA~\citep{tsunemi,slane}, XMM-Newton~\citep{iyudin} 
and  Chandra~\citep{bamba}
demonstrated the non-thermal nature of the X-ray
emission. In this framework, different models interpreting the X-ray spectrum 
yield absorbing column densities for RX~J0852.0-4622 significantly larger
than that of the Vela SNR. Moreover, the
significance level of the $^{44}$Ti yield was later questioned in
a reanalysis of COMPTEL data~\citep{schonfelder}. The Sc-K emission
at about 4~keV expected from the $^{44}$Ti yield is also
controversial: evidence for this line was first claimed from ASCA
SIS0 data~\citep{tsunemi}, but not confirmed by SIS1 data from which
only an upper limit could be deduced~\citep{slane}, whereas a
detection at the 4~$\sigma$ level was obtained from XMM-Newton 
data~\citep{iyudin}. The doubt on the detection of the $^{44}$Ca and 
$^{44}$Sc lines thus affects the interpretation of  RX~J0852.0-4622 as a 
young and close-by supernova remnant.\\

In contrast, on the basis of the absorbing column density deduced from the X-ray
spectrum,~\cite{slane} argue 
in favor of a distance much larger than 200~pc, with the restriction that the remnant be
in front of the Vela Molecular Ridge. Otherwise, this 
concentration of giant molecular clouds, revealed by CO data and
located at a distance of $\sim$ 1--2~kpc, should produce significant
absorption in X-rays in the eastern rim of RX~J0852.0-4622 at a level 
which is not observed. New estimates of the age and distance of this source 
were recently proposed by~\cite{bamba}, on the basis of
the observation of very thin hard X-ray filaments in the
north-western edge with the high angular resolution of the Chandra
satellite. Using an empirical relation~\citep{bamba0} between the
filament width on the downstream side of the shock $w_d$, 
the roll-off frequency
$\nu_{\rm roll}$ of the synchrotron spectrum and the SNR age, the
authors derive an age in the range of $420-1400$~yr and a distance of
$0.26-0.50$~kpc. \\

The possible presence of a compact remnant of the
supernova explosion at the center of RX~J0852.0-4622, first
suggested from ROSAT observations~\citep{aschen98}, was confirmed by
Beppo-SAX~\citep{mereghetti} and Chandra~\citep{pavlov,kargaltsev}; 
if this object is interpreted as a neutron star, as proposed 
by~\cite{chen},
RX~J0852.0-4622 would be the remnant of a core-collapse supernova.
The absorbing column density obtained from the spectrum of this
central object is also significantly higher~\citep{kargaltsev} than
those measured for the Vela remnant, supporting larger distances as
suggested by~\cite{slane}. However, \cite{reynoso} recently interpreted the radio counterpart 
of the central object as due to a planetary nebula; therefore, RX~J0852.0-4622
may also be the result of a thermonuclear explosion.\\

To summarize, there remains
a large uncertainty on the distance of RX~J0852.0-4622; this object
could be as close as the Vela SNR ($\sim$ 290~pc) and possibly in
interaction with Vela, or as far as the Vela Molecular ridge ($\sim$
1~kpc). Even the nature of the explosion of the progenitor remains unclear. 
In addition, the superposition of the Vela SNR and RX~J0852.0-4622 
makes the interpretation of radio and X-ray data difficult. On the other hand,
due to its very weak radio-emission~\citep{combi,duncan} and 
to the non-thermal nature of its X-ray spectrum,
RX~J0852.0-4622 shows remarkable similarities with RX~J1713-3946, 
also detected in the very-high-energy range.\\

This article is organized as follows. In section 2, the main
characteristics of the H.E.S.S. telescope array are reviewed and the 
RX~J0852.0-4622 data set is described. Section 3 is devoted to the analysis
method (gamma-ray selection, angular resolution and spectral
resolution). Results on the gamma-ray morphology of the source are
given in section 4, whereas section 5 is concerned with the spectral
analysis. Section 6 reviews and summarises observations relevant to the 
multiwavelength study of this object, in particular from the X-ray and 
radio bands. Section 7 attempts to derive some general constraints on 
the energetics and emission mechanisms in this source, sections 8 and 9 
then discuss the electronic and hadronic scenarios respectively and 
finally in section 10 we draw some general conclusions.

\section{H.E.S.S. Observations}
H.E.S.S. is an array of four 13 m diameter imaging Cherenkov telescopes 
located in the Khomas Highlands in Namibia, 1800~m above sea level~\citep{HESS}. 
Each telescope has a tesselated mirror with an area of 107~m$^{2}$~\citep{HESSOptics}
and is equipped with a camera comprising 960 photomultipliers~\citep{HESSCamera} 
covering a field of view of 5$^{\circ}$ diameter. During the observations, an array level 
hardware trigger requires each shower to be observed by at least two telescopes within a coincidence window of 60~ns~\citep{HESSTrigger}. Due to the 
efficient rejection of hadronic showers provided 
by stereoscopy, the complete system (operational since December 2003) 
can detect point sources at flux levels of about 1\% of 
the Crab nebula flux near zenith with a significance of 5~$\sigma$ in 25 hours of observation. 
This high sensitivity, the angular resolution of a few arc minutes and the large field of 
view make H.E.S.S. ideally suited for 
the study of the gamma-ray morphology of extended sources.

The region of the supernova remnant RX~J0852.0-4622 was observed with 
the complete H.E.S.S. array between December 2004 and May 2005 for a total observation 
time of 33 hours of ON-source runs. The data were recorded in runs of typical 28 minute 
duration in the so-called ``wobble mode'', where the source is offset from the center of 
the field of view. The offset angles both in right ascension and declination ($\pm$0.7$^{\circ}$, 
$\pm$1.1$^{\circ}$ and $\pm$1.56$^{\circ}$) were chosen in order to provide a full coverage 
of this widely extended supernova remnant. In order to reduce systematic effects due to varying 
observational conditions, quality selection criteria were applied on a run-by-run basis resulting in a total of 20 hours of high-quality data at zenith angles between 20$^{\circ}$ 
and 50$^{\circ}$ (with an average of 30$^{\circ}$).
The energy threshold of the system increases with the zenith angle: for the observations 
presented here, the average threshold was around 250 GeV.

\section{Analysis technique}
The data were calibrated as described in detail in~\cite{HESSCalib}. In a first stage, 
a standard image cleaning was applied to shower images to remove the contamination due to 
the night sky background. Several independent analysis methods are used within the H.E.S.S. 
Collaboration~\citep{HESSAna} to cross-check all results. The results presented in 
this paper were obtained using a 3D-modeling of the light-emitting region of an 
electromagnetic air shower, a method referred to as ``the 3D-model analysis''~\citep{ProcModel3D}. 
All results were verified and confirmed using the standard H.E.S.S. analysis described in 
detail in~\cite{HESSHillas}. We briefly recall the main characteristics of these methods:
\begin{itemize}
\item The standard stereoscopic analysis is based on the Hillas parameters of shower 
images~\citep{HESSHillas}. The incident direction as well as the shower 
impact on the ground are reconstructed from the image axes, whereas parameters directly 
related to the widths and lengths of the images (mean reduced scaled width and mean 
reduced scaled length) are used for gamma-hadron separation. The gamma-ray energy 
is estimated from the image intensity taking into account the reconstructed 
shower geometry, in particular the impact distance. The performance of this 
analysis method as applied to extended sources can be found in~\cite{HESSRXJ}.
\item In the 3D-model analysis, the Cherenkov light emitted by a shower is modeled in 
the following way: the photon origins (photosphere) are distributed according to a 3D-Gaussian law and their directions are assumed to follow an anisotropic angular distribution, with the 
overall requirement of rotational symmetry characteristic of an electromagnetic shower.
The expected number of Cherenkov photons collected by each phototube of a given telescope is 
then calculated by integrating the light yield over the corresponding line of sight. 
A comparison of the observed images to the expected ones by means of a maximum likelihood 
method allows to reject a large fraction of hadronic showers, namely those which are 
not compatible with rotational symmetry. An additional discrimination between gamma-rays and hadrons is provided 
by the lateral spread of the photosphere (or 3D-width) obtained from the likelihood fit. In practice, we use a dimensionless parameter directly related to this quantity, the ``reduced 3D-width'', whose distribution is almost zenith angle independent. 
The energy $E_0$ of the primary gamma-ray is then reconstructed calorimetrically from the number of Cherenkov photons obtained from the fit. A complete 
review of the performance of this analysis method is given in~\cite{HESSModel3D}.
\end{itemize}

\section{Morphology}
\subsection{Background subtraction methods}
For the generation of the excess skymaps for RX~J0852.0-4622, two different methods 
of background subtraction have been applied. The first one is classic: the background 
level is estimated from OFF-source runs, observing sky regions without any gamma-ray 
sources in the field of view. For this purpose, 20 hours of OFF runs taken at similar 
zenith angles are used. All events passing the gamma-ray cuts of the different 
analysis methods, i.e. gamma-ray like background events, are used to estimate 
the background. The second method of background subtraction is more recent and is 
called the ``Weighting Method''~\citep{WeightingMethod}. In this method, the signal and 
the background are estimated simultaneously in the same portion of the sky. In each sky bin 
(treated independently), the signal and the background are estimated from those events 
originating from this bin exclusively. 
%this is done by means of a likelihood fit in which 
Each event is characterized by a discriminating parameter, the reduced 3D-width defined above. 
Since its distribution is fairly different for gamma-rays and hadrons, the respective numbers in 
each population are derived by a likelihood fit.
%the number of gamma-rays is directly obtained as a sum of terms, called ``gamma weights''; similarly, the number of hadrons is obtained as a sum of ``hadron weights''. 
Therefore, no cut on the reduced 3D-width is necessary. The advantage 
of these background subtraction methods is that no assumption on the gamma-ray content in the field of view is necessary. The bin size used in this analysis is 0.05$^\circ$ 
$\times$ 0.05$^\circ$. The images were further smoothed by a Gaussian distribution with a standard deviation of 0.06$^\circ$ to reduce the statistical fluctuations. This procedure was chosen in order to match the H.E.S.S. angular resolution and the binning of the images. The resulting excess maps are in units of integrated excess counts per Gaussian sigma of the smoothing function. 

\subsection{Overall morphology of the remnant}
\label{label:morpho}
\begin{figure}[htbp]
\epsscale{.50}
\plotone{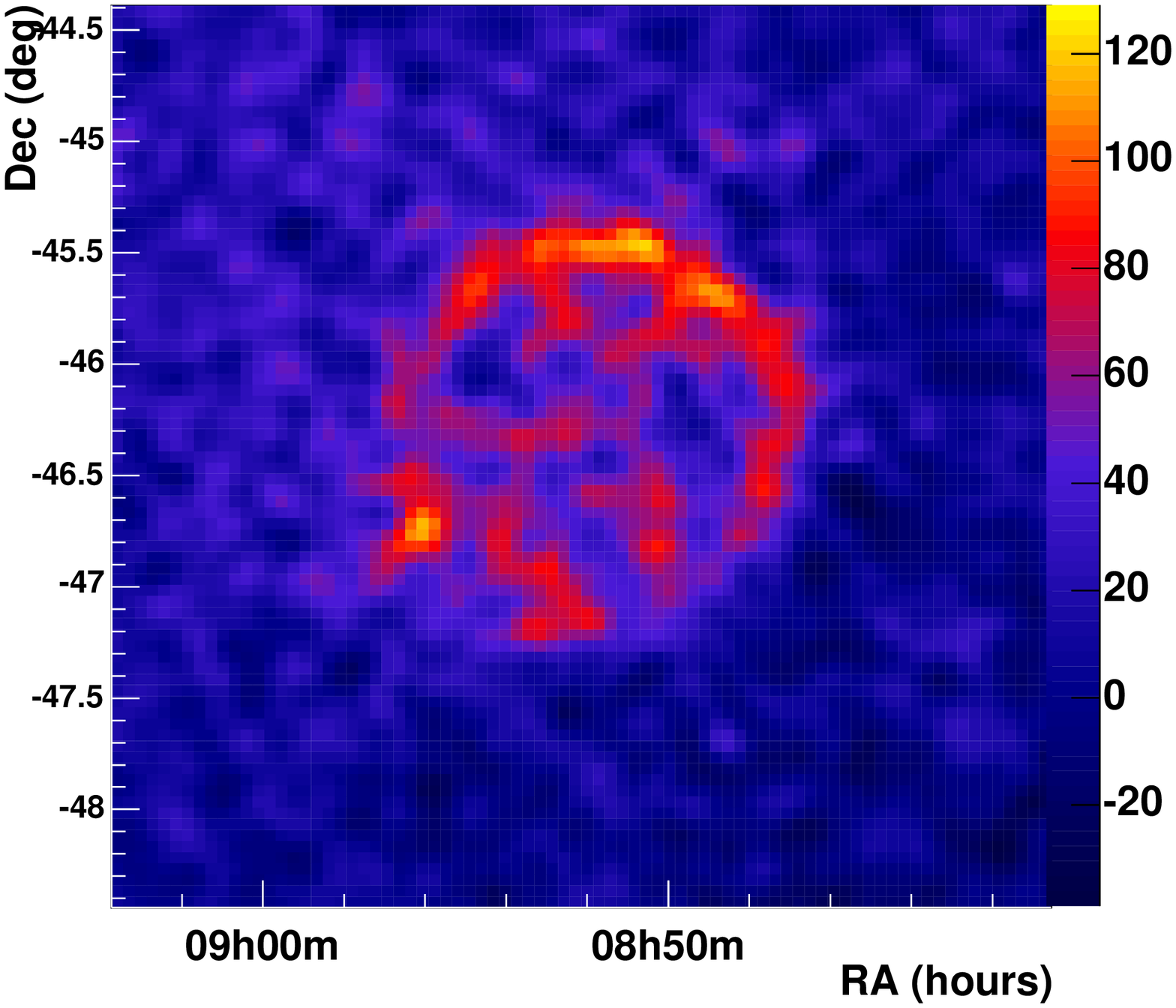}
\caption{Gamma-ray image of RX~J0852.0-4622 smoothed by a 0.06$^\circ$ Gaussian. 
Only events triggering 3 and 4 telescopes were accepted in this analysis leading to a 
better angular resolution, as explained in section~\ref{label:morpho}. The linear colour scale is in units of excess counts per bin.
\label{fig::velajr_34tels}}
\end{figure}

In the study of the morphology of an extended 
source, one of the major objectives is to have the best possible angular resolution. 
In an array such as H.E.S.S. including 4 telescopes placed in a square formation, events triggering 4 telescopes are concentrated in the central region of the array, 
whereas events triggering 2 telescopes, being peripheral, are not so accurately 
reconstructed as the central ones. Therefore, to obtain a high angular resolution 
(of the order of 0.06$^\circ$), one can restrict the analysis to events triggering at least 3 telescopes, which also further reduces the hadronic background. 
The excess skymap in Figure~\ref{fig::velajr_34tels} shows the gamma-ray image of 
RX~J0852.0-4622 in a $4^{\circ} \times 4^{\circ}$ field of view, obtained with the 3D-Model and the Weighting Method by keeping only 3 and 4 telescope events. The gamma-ray content in this skymap presents a much higher statistics than the one obtained with the H.E.S.S. dataset from February 2004~\citep{HESSVelaJr}. The significance is about $19 \, \sigma$ with an excess of $\sim 5200$ events, keeping all events inside a radius of $1^{\circ}$ around the center of the remnant. In order to explore the robustness of the result, the data set was analyzed using the same calibration and analysis method but applying different sets of cuts, which resulted in different resolutions and statistics, all results being compatible with each other. Additionally, the morphology 
was cross-checked using the standard analysis method for the reconstruction and the 
ON-OFF method for the background subtraction. The comparison of the results obtained by 
the 2 methods in a region of $1.2^{\circ}$ radius around the center of the supernova remnant yields a correlation coefficient of $80 \pm 1\%$. These tests show that the gamma-ray morphology of 
the remnant remains consistent when analyzed with different 
sets of cuts or with different background subtraction methods. \\

\begin{figure}[htbp]
\epsscale{.70}
\plotone{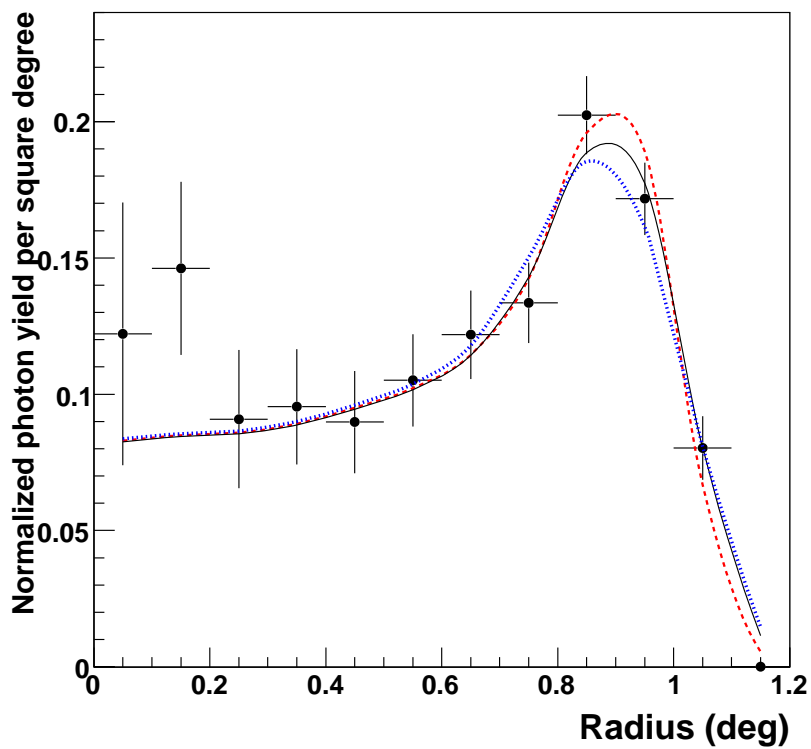}
\caption{Radial profiles around the center of the supernova remnant ($\alpha_{J2000}$ = 8$^h$52$^m$, $\delta_{J2000}$ = $-46^\circ$22') expected for a shell of varying thickness and uniform emission compared to the H.E.S.S. data (black crosses) for the northern part 
of the remnant. The dotted blue 
radial profile has been obtained with a thickness of 22.5$\%$ of the radius of the remnant, 
the dashed red with 12.5$\%$ and the black line with 18.3$\%$. All these histograms have been normalized so that the sum of the contents between 0.3$^{\circ}$ and 1.2$^{\circ}$ is equal to unity.
\label{fig::radprof_toymodel}}
\end{figure}

The morphology appearing from the excess skymap in Figure~\ref{fig::velajr_34tels} reveals a very thin shell of 
1$^\circ$ radius and $\sim$0.2$^\circ$ thickness. It shows several bright regions 
in the north, north-western and south-eastern parts of the supernova remnant. Another feature 
is the remarkably circular general shape of this shell, even if the 
southern part shows a more broken (non-uniform) structure than the northern one. 
This regular morphology resembles very much the image that one would expect from a 
homogeneously emitting shell. In order to investigate the projection effect of the 3D-source 
into a 2D-skymap, a simple geometrical model (``toy model'') of a thin and homogeneous 
shell has been used and adapted to the data. After calculating the projection, 
the obtained skymap is smoothed according to the average point spread function in this 
dataset. The radial 
profiles (i.e the number of events per unit solid angle 
as a function of the distance to the center of the remnant) obtained 
with the ``toy-model'' for different values of the shell thickness are 
then fitted to those obtained from H.E.S.S. data (restricting to 3 and 
4 telescopes events) in the northern part of the remnant (declination 
higher than $-46.3^\circ$) which 
exhibits a clear shell as seen on the gamma-ray excess map. For each value of the shell thickness, the only 
parameter of the fit is the outer radius of the shell. 
Figure~\ref{fig::radprof_toymodel} shows that the bright shell observed 
by H.E.S.S. is well reproduced by the simple geometrical model. 
In Figure~\ref{fig::radprof_toymodel}, the remarkable point to note is 
the small value of the shell thickness giving the best fit: it is equal 
to $18.3\%$ of the radius of the remnant and between 12.5$\%$ 
and 22.5$\%$ at 95\% confidence level. This contrasts with RX~J1713.7-3946, 
another shell-type supernova remnant resolved by H.E.S.S., 
in which the shell thickness that best suited the data was about 45$\%$ of the radius of 
the remnant. This good match of the toy model and the H.E.S.S. data clearly 
shows that the gamma-ray emission detected comes from a thin shell. 

\subsection{Energy dependence of the morphology}
\begin{figure}[htbp]
\epsscale{.50}
\plotone{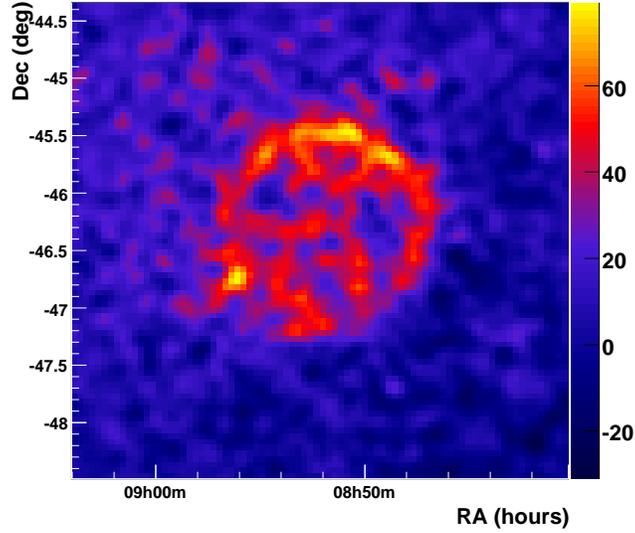}
\caption{Gamma-ray image of RX~J0852.0-4622 at energies lower than 0.5 TeV, smoothed by a 0.06$^\circ$ 
Gaussian. Only events triggering at least 
3 telescopes were kept. The linear colour scale is in units of excess counts per bin.
\label{fig::velajr_inf500GeV}}   
\end{figure}

\begin{figure}[htbp]
\epsscale{.50}
\plotone{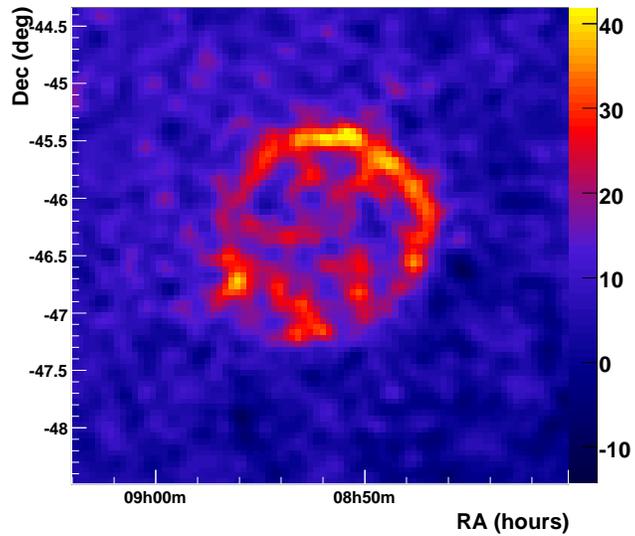}
\caption{Same as figure~\ref{fig::velajr_inf500GeV} for energies higher than 0.5 TeV.
\label{fig::velajr_sup500GeV}}   
\end{figure}

\begin{figure}[htbp]
\epsscale{.70}
\plotone{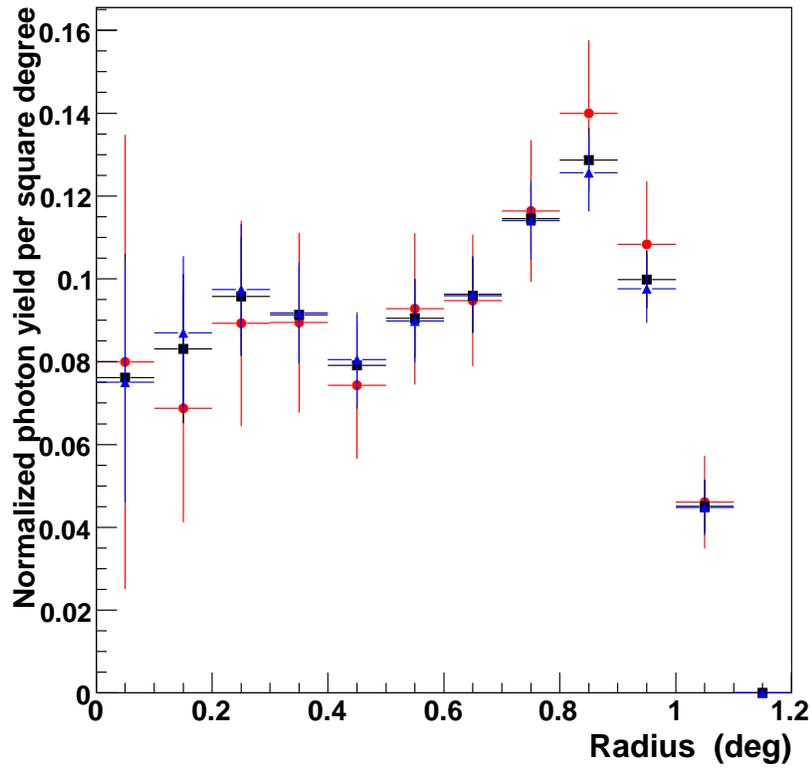}
\caption{Comparison of the radial profiles of the whole remnant in different energy bands. The black 
squares show the radial profile for all energy events, the circles for energies higher 
than 0.5 TeV and the triangles for energies lower than 0.5 TeV. The different distributions 
have all been normalised to unity to enable a direct comparison.
\label{fig::energyband_radprof}}   
\end{figure}

Figures~\ref{fig::velajr_inf500GeV} and \ref{fig::velajr_sup500GeV} show the 
morphology of RX~J0852.0-4622 in two distinct energy bands, E$<$0.5 TeV and E$>$0.5 TeV, keeping only events triggering at least 3 telescopes. The two energy bands were chosen in order to have approximately the same statistics in both datasets. 
Clearly, the morphology of the remnant is the same in the two energy bands. 
The overall radial profile in the two energy bands shown in Figure~\ref{fig::energyband_radprof} 
confirms that the morphology does not vary significantly with energy.

\section{Spectral analysis}
For the spectral analysis, the source region (ON region) is defined by a circle of 1$^\circ$ 
radius centered on the position ($\alpha_{J2000}$ = 8$^h$52$^m$, $\delta_{J2000}$ = $-46^\circ$22'). 
In the study of a point-like source, the restricted selection of events due to the knowledge of the exact gamma-ray origin and the reduced offsets of the source from the center of the camera 
improve the energy resolution. However, in the present case, these two characteristics are lost, 
which results in an average energy resolution $\Delta$E/E $\sim$ 25$\%$ for the 3D-Model, slightly 
larger than for a point-like source. The energy resolution can be improved by 
restricting to 3 and 4 telescope 
events at the expense of a smaller statistics (but with the same statistical significance). 
In this study, the two possibilities (restricting or not to 3 and 4 telescope 
events) were used in order to verify the stability of our results.\\
The spectral analysis requires selection criteria slightly different from those of 
the morphological study. Two cuts were applied independently of the telescope 
multiplicity: a cut on the reduced 3D-width and a cut on the image size at 80 
photoelectrons. All events passing the cuts and with reconstructed direction within 
a region of 1$^\circ$ radius from the center of the 
remnant are considered as ON events.\\ 
For the background estimation, OFF events were 
selected from the same field of view and in the same runs as the ON events by selecting 
the region symmetric to the ON region with respect to the camera center 
(``mirrored background''). A minimum distance of 0.1$^\circ$ between the two regions is required 
in order to avoid any gamma-ray contamination in the OFF data. 
This approach ensures that background events are taken 
at the same zenith angle and offset angles as the ON events, which is crucial because of 
the strong dependence of the effective area upon these two quantities.\\
The energy spectrum of the gamma-ray excess is then obtained by the method of ~\cite{Piron}. 
In this method, an a priori spectral shape is assumed, whose parameters are obtained by fitting the expected distribution to the reconstructed energy distribution. 
In this procedure, gamma-ray acceptances and resolutions calculated from 
simulations are taken into account. \\
 
\begin{figure}[htbp]
\epsscale{.70}
\plotone{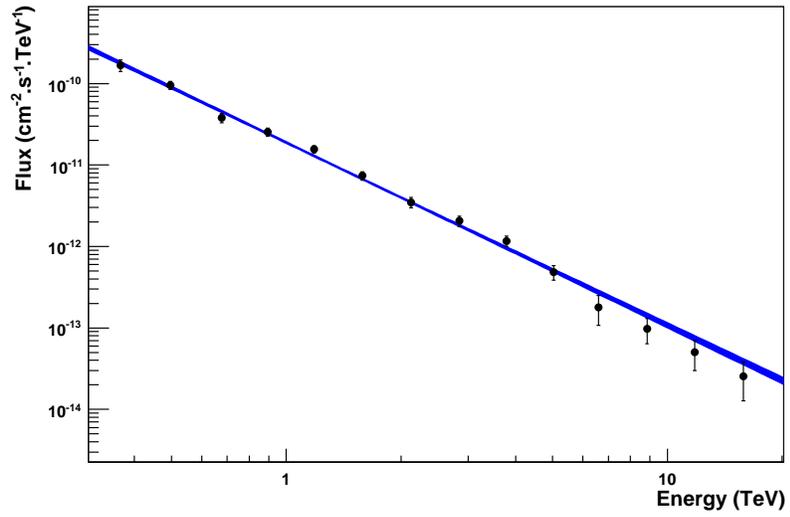}
\caption{Differential energy spectrum of RX~J0852.0-4622, for the whole region of 
the SNR. The shaded area gives the 1$\sigma$ confidence region for the spectral 
shape under the assumption of a power law. The spectrum ranges from 300 GeV to 20 TeV.
\label{fig::velajr_spec_pl}}   
\end{figure}

\begin{figure}[htbp]
\epsscale{.80}
\plotone{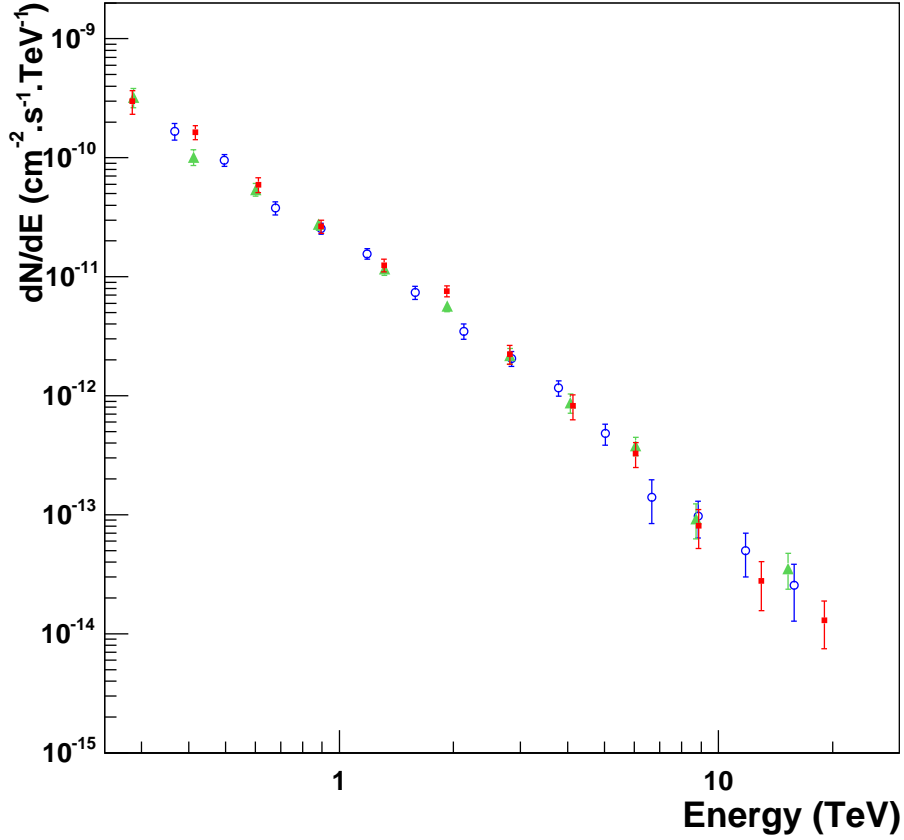}
\caption{Shown are three spectra that were produced to investigate the systematic 
uncertainties. The blue open circles show the spectrum obtained with the 3D-Model 
analysis and using the reflection of the ON region from the center of the camera 
to estimate the background. The red squares spectrum was obtained with the standard 
analysis, the same background estimation but another spectral analysis technique~\citep{HESSRXJ}. 
The green triangles spectrum was obtained by using the standard analysis and the 
same spectral analysis technique as the red spectrum but the background estimation 
was done with OFF runs.  
\label{fig::velajr_allana}}
\end{figure}

The differential energy spectrum thus obtained is shown in 
Figure~\ref{fig::velajr_spec_pl}. It extends from 300 GeV up to 20 TeV.
The spectral parameters were obtained from a maximum likelihood fit of a power law 
hypothesis dN/dE = $\mathrm{N_0}$~$\mathrm{(E/1 \, TeV)^{-\Gamma}}$ to the data, resulting 
in an integral flux above 1 TeV of ($15.2 \pm 0.7_{\mathrm{stat}} \pm 3.20_{\mathrm{syst}}$) $\times$ $10^{-12} \mathrm{cm^{-2}} 
\mathrm{s^{-1}}$ and a spectral index of 2.24 $\pm 0.04_{\mathrm{ stat}} 
\pm 0.15_{\mathrm{ syst}}$. The present data include much more statistics especially at high energy than 
the previous H.E.S.S. measurement~\citep{HESSVelaJr} and a slight 
deviation from a pure power law can be seen in Figure~\ref{fig::velajr_spec_pl}. 
This can explain that the average spectral index is slightly higher than the one from 
the previous measurement: $2.1 \pm 0.1_{\mathrm{ stat}} 
\pm 0.2_{\mathrm{ syst}}$. 
To confirm this spectrum, two other estimates of the spectrum of the whole remnant 
were obtained by using other techniques. These checks ensure that the 
systematics introduced by the spectral analysis technique or by the background 
estimation are small. For these tests, the standard analysis method was applied by 
using a cut on the ``scaled parameters'' and on the image size at 80 photoelectrons. 
In Figure~\ref{fig::velajr_allana}, two different estimates of the spectrum 
are superimposed to the one obtained with the 3D-Model. One was computed by using 
the same background estimation with a mirror region and the other by using a background 
estimation based on OFF runs taken at similar zenith angles. All spectra are compatible 
with each other and all show an indication of deviation from a power-law at high energy.

\section{RX~J0852.0-4622 at other wavelengths}

\subsection{Analysis of ASCA data}
\label{lab:asca}
\begin{figure}[htbp]
\epsscale{.70}
\plotone{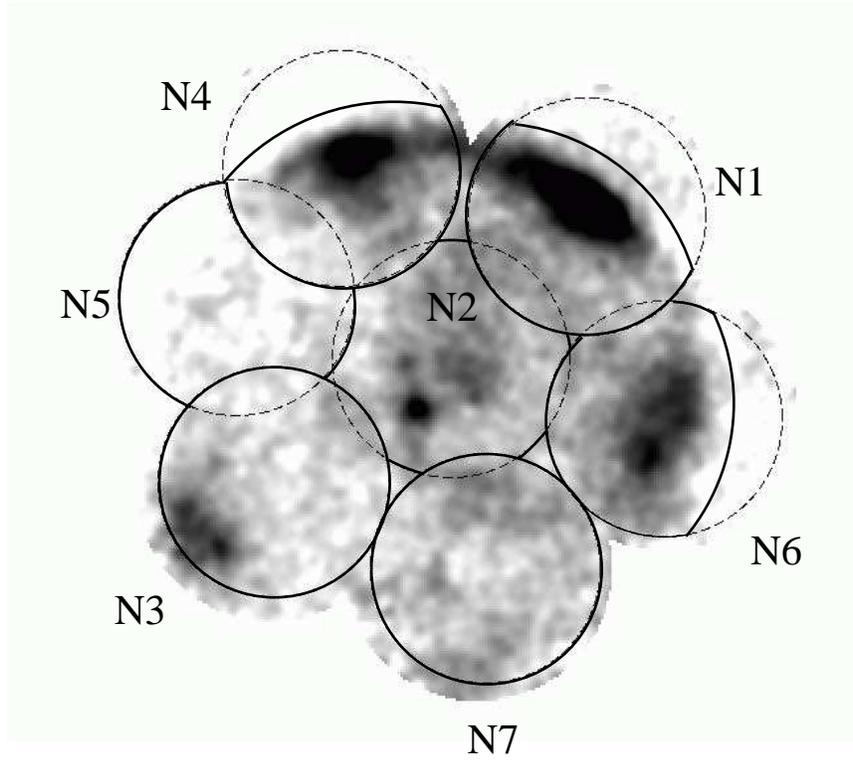}
\caption{ASCA GIS mosaic image in the 0.7-10\,keV energy band~\citep{tsunemi}. The gray-scale is logarithmic. The seven regions used for the
spectral analysis are denoted by solid lines, while the dashed
circles represent the seven pointings.}
\label{fig:asca_fov}
\end{figure}

\begin{figure}
\epsscale{.70}
\rotatebox{-90}{
\plotone{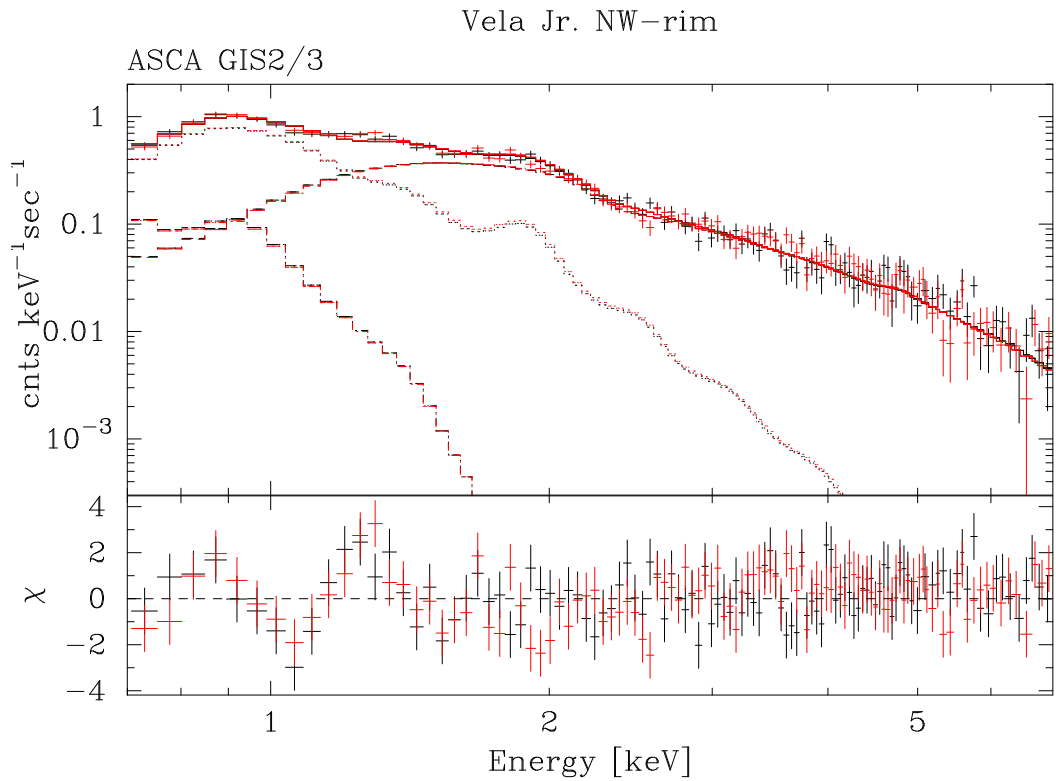}}
\caption{ASCA GIS~(GIS2 and GIS3) spectrum from N1 source region depicted in Figure~\ref{fig:asca_fov}. The different components of the best fit models (absorbed power-law and both thermal spectra) are indicated by the lower histograms. Residuals are presented in the bottom plot.}
\label{fig:asca_specn1}
\end{figure}

ASCA archival data of RX~J0852.0-4622 were used to study the non-thermal emission of the remnant in the X-ray band. Figure~\ref{fig:asca_fov} shows the ASCA GIS~(GIS2 and 3) mosaic image of RX~J0852.0-4622 in the $0.7$--$10.0$~keV energy band~\citep{tsunemi} obtained by using seven distinct pointings~(N1--N7) which almost cover the entire remnant. Standard quality criteria (screening procedures) were applied to ASCA GIS data, and spectra were extracted from the seven non-overlapping regions shown in Figure~\ref{fig:asca_fov}. Since the soft thermal emission from the Vela SNR foreground is spatially variable, background spectra were produced by using blank sky event files, which are considered to contain both non X-ray background and cosmic X-ray background, except for the central pointing~(N2). For this region, due to different observation conditions, the background spectrum was extracted from Large Sky Survey~(LSS) data observed during the ASCA PV phase~\citep{ueda99}. To derive the spectral parameters, the emission was modeled with two components. A thermal model was used to account for the soft thermal emission from the Vela SNR, with a column density fixed at $N_{\mathrm H} = 10^{20} \, {\rm cm^{-2}}$~\citep{lu00} and a temperature of $kT_{\mathrm e}$(low) $= 0.1$~keV, typical values for Vela\footnote{One should note that the GIS response is not well suited to determine the soft component which led us to fix the values of $kT_{\rm e}$(low) and $N_{\rm H}$(Vela).}. An absorbed power law was used for the hard emission. A simultaneous spectral fitting was performed with both GIS instruments from $0.7$ up to $7.0$\,keV. Figure~\ref{fig:asca_specn1} shows the spectrum extracted from region N1 with the best-fit models; it exhibits some line-like feature below $2$~keV, which originates from thermal emission from the Vela SNR and/or RX~J0852.0-4622. The resulting spectral parameters of the fit are the photon index, $\Gamma=2.79\pm 0.09$, the column density for the non-thermal component, $N_{\mathrm H}=6.2^{+1.4}_{-1.3}\times 10^{21}{\rm cm^{-2}}$ and the high temperature of the Vela SNR thermal component, $kT_{\mathrm e}$(high)=0.56$\pm0.1$\,keV. These values are well consistent with previous results~(Slane et al. 2001; Iyudin et al. 2005; Bamba et al. 2005). The derived non-thermal flux from region N1 in the 2--10\,keV energy band is 1.87$\pm0.06 \times 10^{-11} {\rm erg\,s^{-1} cm^{-2}}$ while the thermal flux is $\sim$30 times smaller than the non-thermal flux in 2--10\,keV band. Some residuals are still visible around 1~keV and suggest that another thermal component, which might originate from RX~J0852.0-4622, is needed. For the other six regions, spectra have been fitted with the same procedure as described above. The resultant photon indices are between 2.5 and 2.8 and the total flux for the seven regions is 8.3$\pm0.2 \times 10^{-11} {\rm erg\,s^{-1} \, cm^{-2}}$ ($2$--$10$~keV). All errors described in this section are statistical and given at 90\% confidence level. The systematic error on the flux implied by the procedure has been estimated as follows: the error due to the uncertainty in the instrumental response is $\pm$10\% and the one due to the uncertainty on the surface brightness variation in RX~J0852.0-4622 is $\pm$20\%. 

\subsection{Morphological comparison between H.E.S.S. and X-rays}
\begin{figure}[htbp]
\epsscale{.70}
\plotone{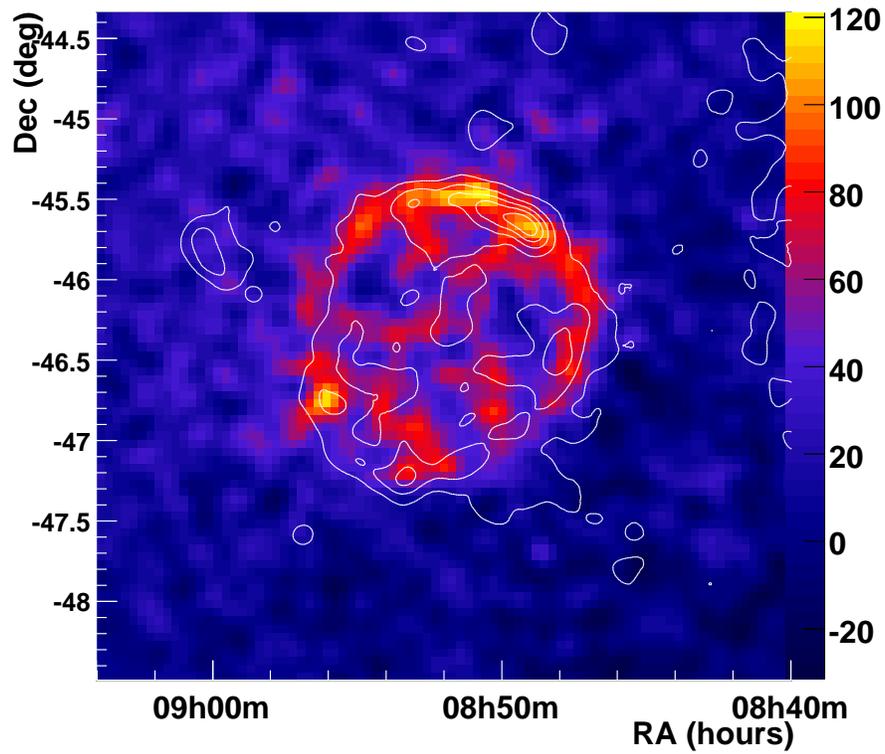}
\caption{Excess skymap of RX~J0852.0-4622 smoothed with a Gaussian of 0.06$^\circ$ 
standard deviation. The white lines are the contours of the X-ray data from the 
ROSAT All Sky Survey for energies higher than 1.3 keV (smoothed with a Gaussian 
of 0.06$^\circ$ standard deviation to enable direct comparison of the two images). 
The linear colour scale is in units of excess counts per bin.
\label{fig::velajr_rosat}}
\end{figure}

\begin{figure}[htbp]
\epsscale{.50}
\plotone{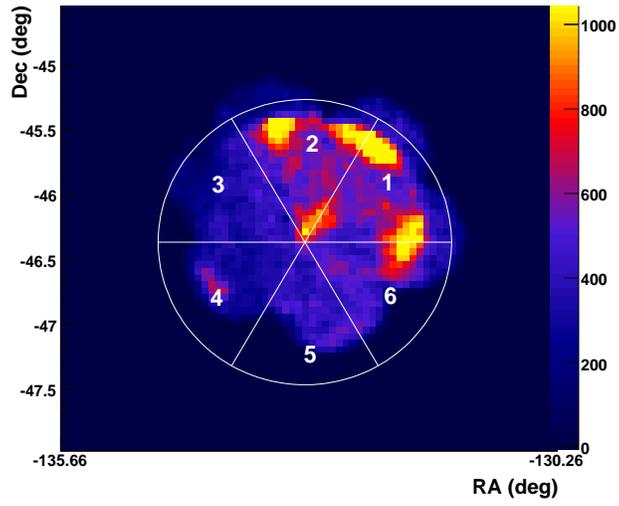}
\caption{ASCA X-ray image of RX~J0852.0-4622. The six regions used in the radial 
profiles are indicated. One can notice that the coverage of the remnant is not 
complete in regions 4, 5 and 6. The linear colour scale is in units of excess counts 
per bin.
\label{fig::asca_camembert}}
\end{figure}

\begin{figure}[htbp]
\epsscale{.99}
\plotone{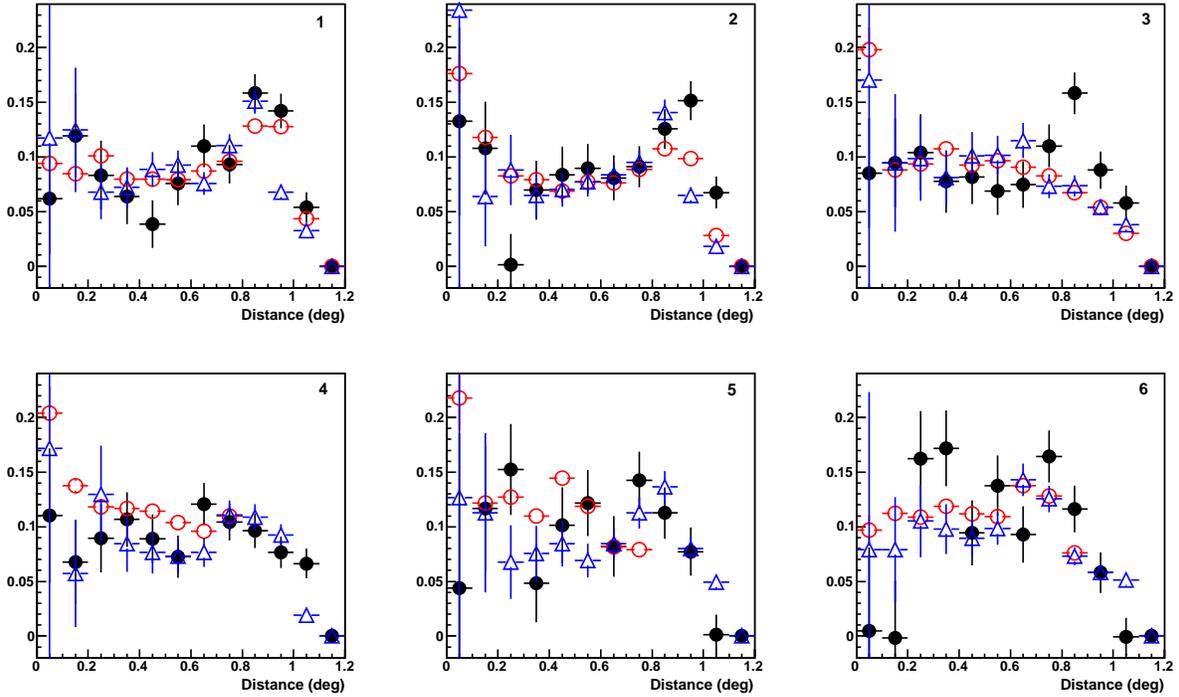}
\caption{Radial profiles for the six regions marked in Figure~\ref{fig::asca_camembert}. 
The black circles represent the H.E.S.S. excess counts per unit solid angle as a 
function of the distance r from the center of the remnant. The open circles represent 
the radial profiles of the ASCA X-ray data. The open triangles represent the radial 
profiles obtained with the X-ray data from the ROSAT All Sky Survey. The different distributions have been normalized to unity in each region to enable a direct comparison. Note that the coverage 
of the SNR by ASCA was not complete in the regions 4, 5 and 6. 
\label{fig::radprof_xray}}
\end{figure}

Figure~\ref{fig::velajr_rosat} presents the gamma-ray excess map obtained by 
H.E.S.S. with the superimposed contours of the X-ray data from the ROSAT All Sky 
Survey. The overall gamma-ray morphology seems to be similar to the one seen 
in the X-ray band, especially in the northern part of the remnant where a brightening is seen in both wavebands. The correlation coefficient between the gamma-ray 
and the X-ray counts in bins of 0.2$^\circ$ $\times$ 0.2$^\circ$ is found 
to be equal to 0.60 and between 0.54 and 0.67 at 95$\%$ confidence level. 
The use of ASCA data gives a 
similar result. The data of the various instruments were then compared to each 
other in six different sectors defined in Figure~\ref{fig::asca_camembert}. 
In each region, the radial profiles were determined. 
A binning of 0.1$^\circ$ for the radial profiles was used; this value, larger than 
the point spread function of each instrument, allows to safely compare their data. 
All the radial profiles were normalized to unity. The results of this study are shown in 
Figure~\ref{fig::radprof_xray}. One should note that, due to an incomplete 
coverage of the SNR in regions 4, 5 and 6, the ASCA data are not reliable 
at distances larger than $\sim 0.7$ degree in these regions and were thus not included; this incomplete coverage 
is visible in Figure~\ref{fig::asca_camembert}. The different radial profiles are in good agreement with each other 
in all regions. Differences between X-rays and TeV gamma-rays seem to appear mainly in 
region 3 where a peak is seen only in the TeV regime. Unfortunately, as the ASCA 
data are incomplete and the ROSAT data are contaminated by the Vela SNR, a more 
quantitative conclusion (for example on the question of the boundaries of the SNR) 
cannot be drawn.

\subsection{Radio observations}
\begin{figure}[htbp]
\epsscale{0.45}
\plotone{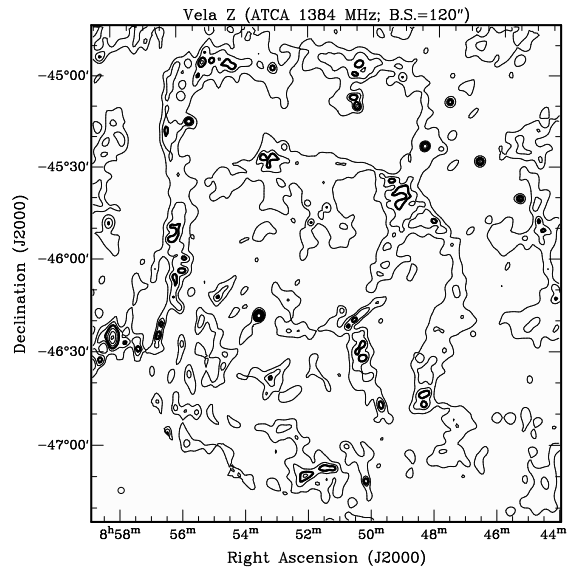}
\caption{ATCA mosaic image of RX~J0852.0-4622 at 1384 MHz. Contours are from 0.01 
to 0.9 in steps of 0.02 Jy/Beam. The synthesized beam of the mosaic ATCA observations is $120$''$\times 120$''. For a better presentation of features inside the 
remnant, RCW 38 is not shown. 
\label{fig::atca}}
\end{figure}

\begin{figure}[htbp]
\epsscale{1}
\plotone{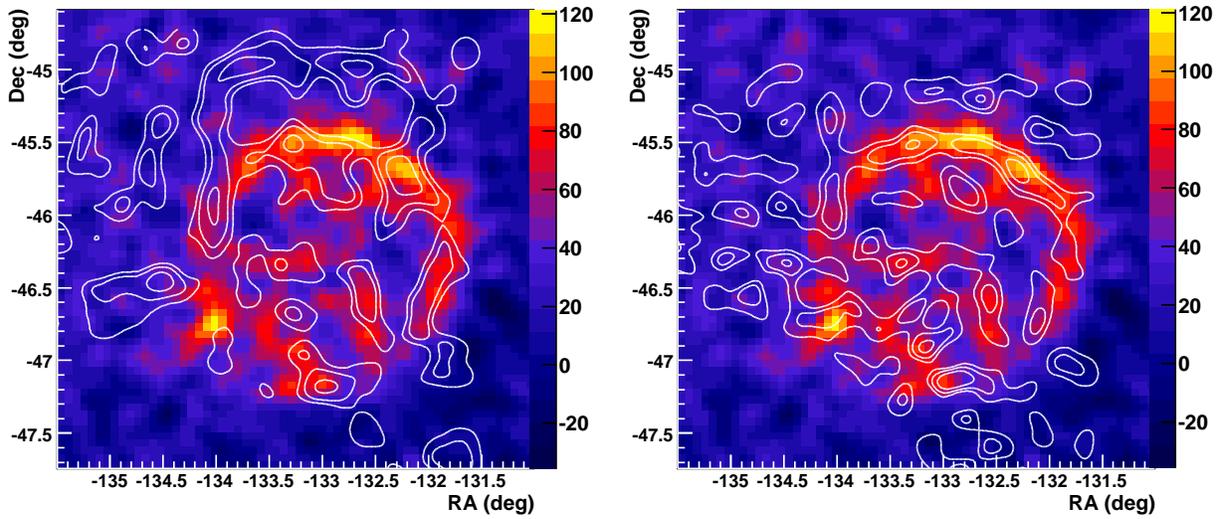}
\caption{Excess skymap of RX~J0852.0-4622 smoothed with a Gaussian of 0.06$^\circ$ 
standard deviation. The white lines are the contours of the ATCA data for a frequency 
of 1384 MHz on the left and 2496 MHz on the right (smoothed with a Gaussian of 0.06$^\circ$ standard deviation to enable 
direct comparison of the two images). The linear colour scale is in units of excess counts per bin.
\label{fig::hess_atca}}
\end{figure}

Mosaic observations of RX~J0852.0-4622 with the Australian Telescope Compact Array 
(ATCA) were undertaken in November 1999 at frequencies of 1384 MHz and 2496 MHz~\citep{Filipovic}. 
Figure~\ref{fig::atca} is a mosaic image of RX~J0852.0-4622 obtained from 110 pointings 
at 1384 MHz. The image suffers from sidelobes originating from the strong radio source 
CTB 31 (RCW 38). A certain similarity between the H.E.S.S. image (Figure~\ref{fig::velajr_34tels}) 
and the radio image of RX~J0852.0-4622 (Figure~\ref{fig::atca}) can be noticed at first glance. 
The overall morphology appears to be similar: many features seen in the TeV skymap coincide 
well with the radio image, such as the bright region in the northern part of the shell 
and the central filamentary structures inside the SNR. Indeed, a high correlation can be 
seen when superimposing the radio contours to the H.E.S.S. images, as shown in 
Figure~\ref{fig::hess_atca}. The ATCA contours directly match 
the structures from the H.E.S.S. skymap and the features obtained inside the remnant are also 
well reproduced. 

\subsection{CO observations}
\begin{figure}[htbp]
\epsscale{0.8}
\plotone{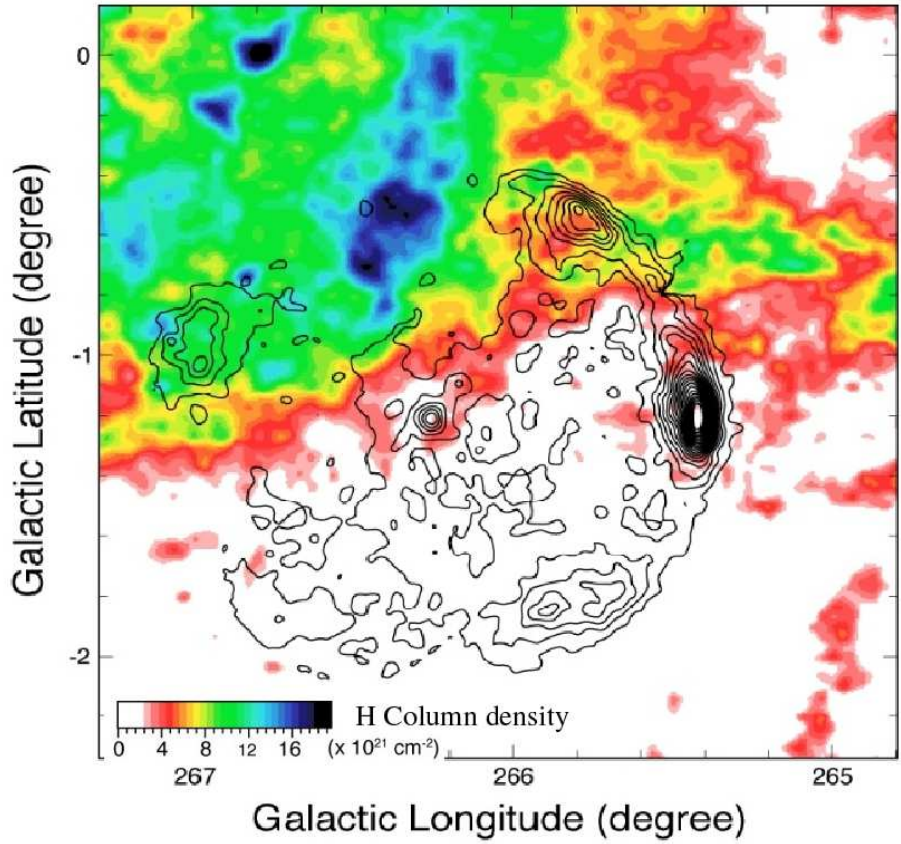}
\caption{Integrated molecular column density, in linear colour scale. Overlaid 
are the contours of the ASCA X-ray excess image. Note that the image is shown in galactic coordinates.
\label{fig::fikui}}
\end{figure}

\begin{figure}[htbp]
\epsscale{0.8}
\plotone{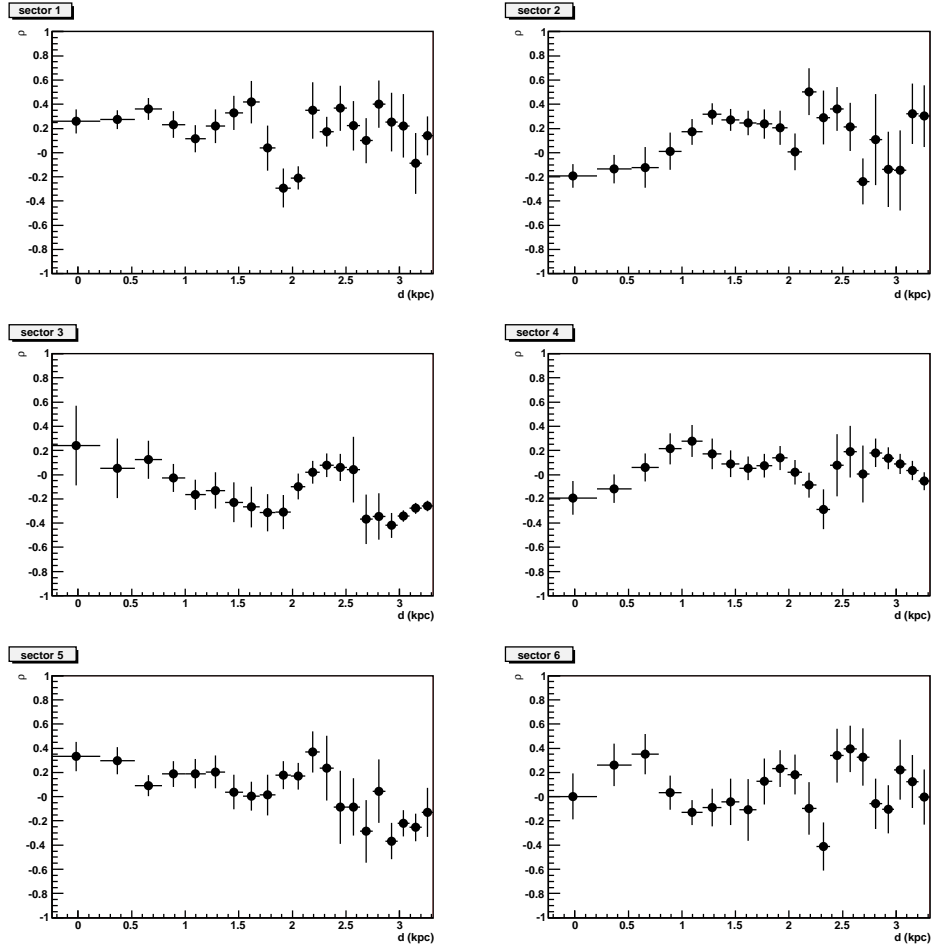}
\caption{Correlation coefficient between the CO intensity and the H.E.S.S. data in 
different distance intervals for each region defined in Figure~\ref{fig::asca_camembert}. No clear correlation appears in any of the regions analysed. 
\label{fig::correl_CO_hess}}
\end{figure}

CO data at 2.6 mm wavelength of the region of the Vela Molecular Ridge and its 
surroundings were taken with the 4-m, mm and sub-mm telescope NANTEN in 1999 ~\citep{moriguchi}. 
Figure~\ref{fig::fikui} shows the integrated molecular column density derived from CO data, in the RX~J0852.0-4622 region. 
A high density is clearly visible in the eastern part 
of the remnant (which corresponds to regions 3 and 4 defined in Figure~\ref{fig::asca_camembert}) 
due to the presence of the Vela Molecular Ridge (VMR). 
Unfortunately, in region 4, uncertainties concerning the X-ray data 
(incomplete coverage and contamination by the Vela SNR foreground) 
preclude any firm conclusion concerning a possible anti-correlation with the CO data. 
On the other hand, no clear evidence of interaction between very high 
energy particles and the VMR is seen in the H.E.S.S. data since the gamma-ray flux does 
not vary by more than a factor of 2 from the eastern to the western sides, whereas the molecular 
column density drops by a factor of $\sim$20. Figure~\ref{fig::correl_CO_hess} shows the 
correlation coefficient between the CO and the H.E.S.S. data calculated in different intervals 
of distance varying from $-0.2$ to 3.2 kpc, in the 6 regions defined previously. 
Distances were estimated by using a galactic rotation model~\citep{brand}. No clear 
correlation can be seen in this Figure~\ref{fig::correl_CO_hess}.

\section{Constraints on source energetics and emission mechanism}
As stated above, the parameters of the supernova remnant RX~J0852.0-4622 are not well known, in particular its age and its distance. Indeed, the remnant could be as close as the Vela SNR ($D \approx 200$~pc) and very young, or as far as the Vela Molecular Ridge ($D \approx 1$~kpc). This leads to a change in the prediction of the X-ray or gamma-ray luminosity by a factor 25 according to the assumed distance. However, beside the distance, other observational characteristics of the remnant provide some helpful constraints.

\subsection{Main constraints beside the spectral analysis}
\label{lab:cons}
The analysis of ASCA data enabled~\cite{slane} to derive a limit on the thermal content of the remnant emission and therefore a limit on the density $n_0$ of the thermally emitting material in the remnant. Using the assumption of thermal equilibrium, the limit obtained is $n_0 < 2.9 \times 10^{-2} (D/ 1 \, \mathrm{kpc})^{-1/2} f^{-1/2} \, \mathrm{cm^{-3}}$, where $f$ is the filling factor of a sphere taken as the emitting volume in the region chosen. It should be noted that this limit is restricted to a gas with temperature above 1~keV because of the contribution of the thermal emission from the Vela SNR at lower energies; higher densities of cooler material are thus not excluded. Furthermore, if the shocks are strongly modified by the accelerated particles, the shock heating is substantially reduced and the X-ray data could be consistent with higher densities. 
Another important piece of information comes from the measured width of the filaments observed by the Chandra satellite. If we assume that these filaments are due to the outer shock, their width determines the downstream magnetic field $B_d$. The thickness of the filament observed by Chandra in the 2-10~keV energy band is: $ w_d = 0.24_{-0.07}^{+0.19} \left( \frac{D}{1 \, {\rm kpc}} \right) \, {\rm pc} $. This led~\cite{bamba} to suggest a value of $B_d \sim 500 \, \mu{\rm G}$ for a distance of 300~pc. The field evaluation based on the work of~\cite{bervolk} gives values $B_d \sim 350 \, \mu{\rm G}$. Such values imply that the magnetic field is highly amplified. 
Finally, the morphological analysis of the H.E.S.S. data sets an upper limit on the thickness of the shell $\Delta R < 22.5$\% of the radius of the supernova remnant. For electrons, which rapidly lose their energy by synchrotron and inverse Compton radiation, the escape time out of the shell into the remnant interior (by diffusion and convection) should be larger than the characteristic time of energy losses. Otherwise, their interactions with photons from the cosmic microwave background would lead to a gamma-ray emission region more extended than that observed by H.E.S.S.. For protons, for which the time scale for energy losses is longer, the escape time must be larger than the age of the supernova remnant. The characteristic escape time of particles is calculated with the formula $t_{\rm esc} = (t_{\rm diff}^{-1} + t_{\rm conv}^{-1})^{-1}$, $t_{\rm diff}$ being the diffusive escape time in the Bohm diffusion regime and $t_{\rm conv}$ the convective escape time. The diffusive escape time is given by $t_{\rm diff} = \frac{\Delta R^2}{2 D_{\rm diff}}$ with $D_{\rm diff}$ the diffusion coefficient for $100$~TeV particles and $\Delta R$ the upper limit on the width of the shell as resolved by H.E.S.S.; the convective escape time is given by $t_{\rm conv} = \frac{4 \Delta R}{V_{\rm shock}}$ with $V_{\rm shock}/4$ the flow velocity into the interior in the shock frame, as derived from~\cite{truelove} and reported in Table~\ref{tab:evol3}. This allows us to calculate a lower limit on the magnetic field $B_{esc}$, reported in Table~\ref{tab:evol3}, in order to confine particles of $100 \, \rm{TeV}$ within the thin shell resolved by H.E.S.S.. However, it should be noted that the shell thickness in the case of a proton model also reflects the thickness of the gas target and is not necessarily a good indicator of thickness of the zone filled by accelerated particles (unlike the electron case where the CMB radiation provides a uniform target). Nevertheless the condition that the acceleration time scale associated with 100 TeV particles be less than the age of the remnant, or the equivalent condition that the associated diffusion length scales be small compared to the shock radius, lead, within factors of order unity, to identical lower limits on the magnetic field strength. 

\begin{table}[tbp]
\begin{center}
\begin{tabular}{|c|c|c|c|c|c|c|c|}\hline
 $D$ & $n$ & $M_{ej}$ & $V_{\rm shock}$ & $B_{esc}$ & $W_p^{\rm tot}$ & efficiency & Age \\ 
  (pc) & ($\mathrm{cm^{-3}}$) & ($\mathcal{M_{\bigodot}}$) & ($\rm km \, s^{-1}$) & ($\mu \mathrm{G}$) & ($10^{49} \, {\rm erg}$) & (\%) & (years) \\ \hline
200 & 0.1 & 1.4 & 6670 & 29.0 & 10 & 10 & 293\\ 
    & & 14 & 3751 & 51.6 & 10 & 10 & 521 \\ \hline
200 & 0.01 & 1.4 & 11862 & 16.3 & 100 & 100 & 165 \\
    & & 14 & 6670 & 29.0 & 100 & 100 & 293 \\ \hline
600 & 0.1 & 1.4 & 2032 & 18.5 & 90 & 90 & 2292 \\
    & & 14 & 1645 & 39.2 & 90 & 90 & 3565 \\ \hline
600 & 0.01 & 1.4 & 5203 & 12.4 & 900 & 900 & 1127 \\
    & & 14 & 2926 & 22.0 & 900 & 900 & 2004 \\ \hline
1000 & 0.1 & 1.4 & 945 & 20.1 & 250 & 250 & 7531 \\
     & & 14 & 944 & 34.5 & 250 & 250 & 9080 \\ \hline
1000 & 0.01 & 1.4 & 2988 & 9.3 & 2500 & 2500 & 2871 \\
     & & 14 & 1994 & 19.4 & 2500 & 2500 & 4900 \\ \hline
\end{tabular} 
\end{center}
\caption{Magnetic field $B_{esc}$, total energy of accelerated protons $W_p^{\rm tot}$, efficiency and age of the supernova remnant (assuming an energy explosion of $10^{51} \, \rm erg$) for different values of distance $D$, density of the medium $n$ and ejected mass $M_{ej}$. For the latter, we have chosen two possible values for RX~J0852.0-4622: $1.4 \, \mathcal{M_{\bigodot}}$ which is typical for SNIa and $14 \, \mathcal{M_{\bigodot}}$ as an average value for SNII. $B_{esc}$ is the lower limit on the magnetic field allowing to confine particles of $100 \, {\rm TeV}$ in the thin shell resolved by H.E.S.S.. It has been calculated from the value of the shock velocity $V_{\rm shock}$ indicated in this table as derived from~\cite{truelove}.}
\label{tab:evol3}
\end{table}  

\subsection{Spectral constraints and modeling of emission processes}
Another constraint comes from the broadband spectral energy distribution (from radio to gamma-rays) as interpreted by a model of emission processes taking place in the supernova remnant. The objective is to constrain parameters like the magnetic field, the density of the medium and the injection spectrum of the primary particles, thanks to a multi-wavelength study.

\subsubsection{The multi-wavelength data}
The gamma-ray spectrum obtained by analyzing the H.E.S.S. data is well described by a power-law with a photon index of $2.24 \pm 0.04_{\mathrm{stat}} \pm 0.15_{\mathrm{ syst}}$:
$$ \phi(E) = \frac{\rm dN}{\rm dE} = (1.90 \pm 0.08_{\mathrm{stat}} \pm 0.40_{\mathrm{ syst}}) \times 10^{-11} \mathrm{cm^{-2} \, s^{-1} \, TeV^{-1}} \left( \frac{E}{1 \mathrm{TeV}} \right)^{-2.24 \pm 0.04 \pm 0.15} $$
This flux can be translated into a global energy flux $\omega_{\gamma}$ between 1 and 10~TeV by using the formula:
$$ \omega_{\gamma}(1-10 \mathrm{TeV}) = \int_{1\,\mathrm{TeV}}^{10\,\mathrm{TeV}} E \times \phi(E) \, dE = 5.4 \times 10^{-11} \mathrm{\, erg \, cm^{-2} \, s^{-1}} $$
which corresponds to a $\gamma$-ray luminosity $L_{\gamma} = 2.6 \times 10^{32} \left(\frac{D}{200 \, \mathrm{pc}} \right)^2 \mathrm{erg \, s^{-1}}$. If one assumes that the gamma-ray flux is entirely due to proton-proton interactions, we can estimate the total energy $W_p$ of accelerated protons in the range $10 - 100$~TeV required to produce this gamma-ray luminosity. In this energy range, the characteristic cooling time of protons through the $\pi^0$ production channel is approximately independent of the energy and can be estimated by~\citep{felixbouquin}: $\tau_{\gamma} = 4.4 \times 10^{15} \left(\frac{n}{1 \, \mathrm{cm^{-3}}} \right)^{-1} \, \rm s$. Thus:$$W_p (10 - 100 \, \mathrm{TeV)} \approx L_{\gamma} \times \tau_{\gamma} \approx 1.1 \times 10^{48} \left( \frac{D}{200 \, \mathrm{pc}} \right)^2  \left( \frac{n}{1 \, \mathrm{cm^{-3}}} \right)^{-1} \mathrm{erg}$$
Assuming that the proton spectrum continues down to $E \approx 1$~GeV with the same spectral slope as that of the photon spectrum, the total energy injected into protons is estimated to be:
$$ W_p^{\rm tot} \approx 10^{49} \left( \frac{D}{200 \, \mathrm{pc}} \right)^2  \left( \frac{n}{1 \, \mathrm{cm^{-3}}} \right)^{-1} \mathrm{erg}$$ 
Values of $W_p^{\rm tot}$ are reported in Table~\ref{tab:evol3} for different distances and densities of the ambient medium.\\
The X-ray spectral analysis of the whole remnant in the 2-10 keV energy band was presented in section~\ref{lab:asca}: the non-thermal spectrum is well described by a power-law with a spectral index of $2.65 \pm 0.15$ and a flux $F_X = 8.3 \pm 0.2 \times 10^{-11} \, \mathrm{erg \, cm^{-2} \, s^{-1}}$.
The fluxes at 1.40~GHz and 2.42~GHz were taken from the analysis of the Parkes data by~\cite{duncan}. Later on, we shall use these multi-wavelength measurements when comparing the H.E.S.S. spectral data to broadband models.

\subsubsection{Modeling the emission processes}
In the simple model used here, we assume that primary particles (protons and electrons) are injected at a constant rate with the same spectral shape, namely a power-law with an exponential cut-off at the energy $E_0$, into a spherical shell of fixed thickness $\Delta R$. As seen previously, the H.E.S.S. data require $\Delta R < 22.5$\% of the remnant radius. The injection is supposed to last a time T (the age of the supernova remnant) in a region of magnetic field $B$ and ambient density $n$. The electron to proton ratio $K_{ep}$ is a free parameter. The energy distribution of the electrons is calculated at a fixed time $t$ by taking into account energy losses due to synchrotron radiation, inverse Compton scattering and bremsstrahlung, while the proton spectrum is calculated by taking into account the escape out of the shell by diffusion (in the Bohm diffusion regime) and convection, as described in section~\ref{lab:cons}. Adiabatic losses are neglected. The broadband spectrum of the source is then derived by taking into account p-p interactions, synchrotron radiation (of primary and of secondary electrons produced via p-p interactions), inverse Compton scattering and bremsstrahlung. Concerning the energy density of the target photons in the Inverse Compton process, we added the contribution of the cosmic microwave background, $0.25 \, {\rm eV \, cm^{-3}}$, and that of the galactic seed photons, namely on average: $0.5 \, {\rm eV \, cm^{-3}}$ for the optical star light and $0.05 \, {\rm eV \, cm^{-3}}$ for the infra-red background \citep{mathis}.\\
%canonical values of the energy density of the seed photons: $0.25 \, {\rm eV \, cm^{-3}}$ for the cosmic microwave back%ground, $0.5 \, {\rm eV \, cm^{-3}}$ for the optical star light and $0.05 \, {\rm eV \, cm^{-3}}$ for the infra-red bac%kground. \\  
It is clear that such a model oversimplifies the acceleration process in an expanding remnant, as discussed by e.g.~\cite{drury2} and~\cite{berezhko}. To this must be added the uncertainties introduced by the dynamics of the ejecta, the nonuniform structure of the ambient medium, and the complexities of the reaction of the accelerated particles on both the magnetic field and the remnant dynamics. However, as a starting point for estimates such a simple model is still, we feel, useful, at least for those cases where the remnant evolution is relatively smooth and the emission is not dominated by relic particles injected and accelerated at earlier times~\citep{yamazaki}.\\

In this study, we explored two different cases of distance (200~pc and 1~kpc), both for the electronic process (gamma-rays mainly produced by inverse Compton scattering) and for the hadronic process (gamma-rays mainly produced by p-p interactions). Using the free expansion and Sedov-Taylor phase equations~\citep{truelove}, one can easily find that, in the nearby case, the supernova remnant should be very young ($\sim 500$~years). On the contrary, in the distant case, the supernova remnant would be rather old ($\sim 5000$ years). In the following scenarios, the total energy injected into the protons is fixed to $10^{50}$~erg (i.e 10\% of the energy of explosion of an average supernova), and the width of the shell must be smaller than 22.5\% of the radius of the remnant. The age of the remnant is assumed to be 500~years at 200~pc and 5000~years at 1~kpc. The other parameters, namely the characteristics of the injection spectrum (spectral index and cut-off energy), the electron/proton ratio $K_{ep}$ at the injection level, the density of the medium and the magnetic field are free parameters in the fit.

\section{The electronic scenario}
If the TeV emission is mainly due to inverse Compton scattering, one should note that, independently of the distance assumed, the ratio of the X-ray flux and the gamma-ray flux determines the value of the magnetic field $B$. In the case of RX~J0852.0-4622, assuming that Inverse Compton emitting electrons are contained in a volume equal to the one responsible for synchrotron emission, one can easily deduce that the magnetic field has to be close to $6 \, \mu{\rm G}$.   

\subsection{The case of a nearby supernova remnant (D = 200~pc)}
Figure~\ref{fig:fitlepton} a) shows the best fit obtained, together with the measurements at different wavelengths: the injection spectrum follows a power-law of index 2.4 and an exponential cut-off at 40~TeV; the value of the magnetic field is $6 \, \mu {\rm G}$ and the density of the medium should be lower than $0.1 \, {\rm cm^{-3}}$ so that the gamma-ray flux produced by p-p interactions should not be significant. One can clearly notice that the radio flux predicted by our model is about 3 times larger than the one observed in the radio range by Parkes. Even more constraining is the thickness of the shell observed in gamma-rays, which is inconsistent with the observations. Indeed, at 200~pc, our limit on the width of the shell implies $\Delta R < 0.78$~pc, which leads to an escape time by diffusion and convection of about 300~years for an ambient density of $0.1 \, {\rm cm^{-3}}$ and an energy of 40~TeV. This value is lower than the age of the remnant ($\sim 500$~years) but also lower than the synchrotron loss time of 8700~years. Electrons above $\sim 20$~TeV will escape the shell and thus automatically produce gamma-rays by inverse Compton scattering on the cosmic microwave background. These gamma-rays have on average an energy $E_{\gamma}$ greater than $500$~GeV. Therefore, we expect to observe a much thicker shell for $E_{\gamma} \ge 500$~GeV. The present analysis of H.E.S.S. data does not show any variation of the morphology of the remnant with the energy, which highly disfavours this scenario.     

\subsection{The case of a distant supernova remnant (D = 1~kpc)}
In the case of a distant object, the magnetic field must also be close to $6 \, \mu {\rm G}$; the only difference comes from the fact that energy losses are no more negligible since the supernova remnant is older ($\sim 5000$~years). These energy losses tend to steepen the electron spectrum and, in order to compensate this effect, the cut-off energy has to be increased in comparison to the preceding case. The parameters of our best fit is an injection spectrum following a power-law with an index of 2.4, a cut-off energy of 80~TeV and an electron/proton ratio $K_{ep} = 3.5 \times 10^{-2}$ (Figure~\ref{fig:fitlepton} b)). The different multi-wavelength data are reasonably reproduced despite a radio flux three times larger than the observational data from Parkes. This last point is not critical since the fit could be improved by including non-linear acceleration effects, which are expected to lead to a steeper rise of the synchrotron SED with frequency above the radio range.\\
In this case, the characteristic time of synchrotron losses is 3700 years for the maximal energy 80~TeV, while the age of the SNR varies from 4000~years to 9000~years depending on the density of the medium in which it evolves. One can easily find that the escape time is either larger than the synchrotron loss time or larger than the age of the remnant, and is thus irrelevant in this case. 

\section{The hadronic scenario}
First, one can see from Table~\ref{tab:evol3} that the only way to explain the entire gamma-ray flux by proton-proton interactions in a homogeneous medium is to assume that RX~J0852.0-4622 is a nearby supernova remnant (D $\leq 600$~pc). Indeed, for larger distances and a typical energy of the supernova explosion, the acceleration efficiency would be excessive (assuming a uniform ambient density compatible with the limit implied by the non-detection of thermal X-rays). Nevertheless, a distance of 1~kpc should also be considered if RX~J0852.0-4622 is assumed to be the result of a core collapse supernova which exploded inside a bubble created by the wind of a massive progenitor star, as proposed by~\cite{bervolk06} for the SNR RX~J1713.7-3946. According to stellar wind theory~\citep{chevalierliang}, the size of the bubble evolves according to the formula: $R = 45 \left(\frac{n_0}{1 \, \rm{cm^{-3}}}\right)^{-0.2}$~pc. For a density of 1~$\mathrm{cm^{-3}}$, the radius of this bubble would be equal to 45~pc. In the case of a close by supernova remnant, its size would be significantly lower than the size of the bubble and the hypothesis of a homogeneous medium would be satisfactory. On the opposite, for larger distances ($D \sim 1$~kpc), the presence of the Vela Molecular Ridge can produce a sudden increase of the density leading to a smaller bubble ($15.6$~pc for a density of $200 \, \mathrm{cm^{-3}}$), which would make the proton-proton interactions efficient at the outer shock.\\
In any case, independently of the distance of the remnant, the extension of the H.E.S.S. spectrum up to 20~TeV implies an energy cut-off $E_0$ higher than 100~TeV. If we assume that energy losses are negligible over the lifetime of the remnant, the synchrotron spectrum can then be approximated by the formula~\citep{reyn}:
$$F(E) \propto E^{-(\Gamma+1)/2} \exp[-(E/E_m)^{1/2}]$$ 
with 
$$E_m \approx 0.02 (B/10 \, \mu G) (E_0/10 \, {\rm TeV})^2 \, {\rm keV}$$
This relation implies that the magnetic field should be lower than $10 \, \mu {\rm G}$ to obtain a synchrotron peak centered at an energy lower than 2~keV. The gamma-ray flux would then be entirely produced by Inverse Compton scattering (as seen previously in the electronic process) which enables to exclude this situation. Therefore, the magnetic field should be high enough to produce significant energy losses during the lifetime of the remnant $t_0$ ($\sim 500$~years) and create a break in the synchrotron spectrum. Refering to~\cite{kifune}, to obtain a break at an energy $E_b$ close to $\sim 0.2$~keV, one would need a magnetic field higher than 40~$\mu \rm G$:
$$ E_b = 2.9 \left(\frac{B}{10 \, \mu \rm G} \right)^{-3} \left(\frac{t_0}{10^3 \, \rm years} \right)^{-2} \rm keV $$ 

\subsection{The case of a nearby supernova remnant (D = 200~pc)}
In this case, our best fit is obtained for an injection spectrum in the form of a power-law with an index of 2.1, a cut-off at 110~TeV and a very low electron/proton ratio of $K_{ep} = 2.4 \times 10^{-6}$. The density of the medium is 0.2~$\rm cm^{-3}$ and the magnetic field amounts to 120~$\mu {\rm G}$: these two values are compatible with both the limit implied by the absence of thermal X-rays and with the thin filaments resolved by Chandra. On the other hand, one can note in Figure~\ref{fig:fithadron} a) that such a model with the above parameters does not provide a good description of the ASCA data, since the energy losses tend to steepen the electron spectrum. However, knowing the difficulty of the X-ray spectral analysis, this point cannot be used to exclude this scenario. Furthermore, a better agreement could be obtained by simply relaxing the assumption used in the model that electrons and protons have similar injection spectra.

\subsection{The case of a distant supernova remnant (D = 1~kpc)}  
In this last case, if the density were low enough as to be compatible 
with the absence of thermal X-rays, Table~\ref{tab:evol3} shows that 
the values of $W_p^{\rm tot}$ required to account for the total observed 
gamma-ray flux exceed the total energy of the supernova explosion 
assumed in the present calculation or would require an anomalously energetic 
explosion. As stated previously, a way out of this difficulty would be to consider the case of a bubble created 
by the wind of the massive progenitor. Actually, our best fit is obtained for an injection spectrum in the form of a power-law with an index $\Gamma = 2.0$, a cut-off energy at $E_0 = 100$~TeV and an electron/proton ratio of $4.5 \times 10^{-4}$ (Figure~\ref{fig:fithadron} b)). The density of the medium is found to be $2.0 \, \rm{cm^{-3}}$, which is acceptable in the framework of the bubble scenario. The magnetic field, $85 \, \mu {\rm G}$, is compatible with the very thin shell resolved by H.E.S.S.. Finally, one should note that our model perfectly reproduces the radio and the H.E.S.S. data, but only approximately the slope coming from the ASCA spectral analysis.

\section{Conclusions}
\begin{figure}[htbp]
\begin{minipage}{0.47\linewidth}
\centering
\includegraphics[width=\linewidth]{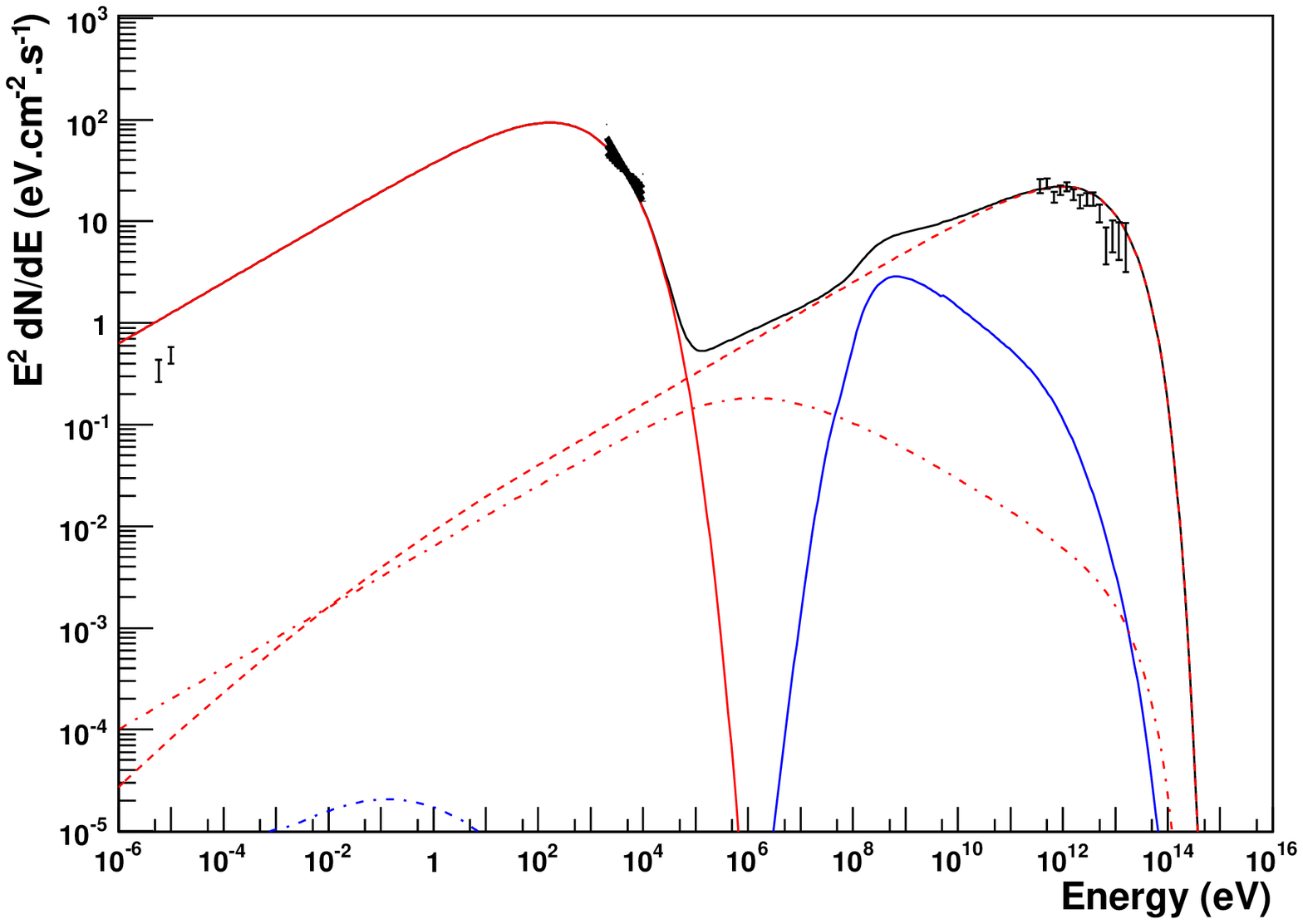}
\end{minipage}
\hfill
\begin{minipage}{0.47\linewidth}
\centering
\includegraphics[width=\linewidth]{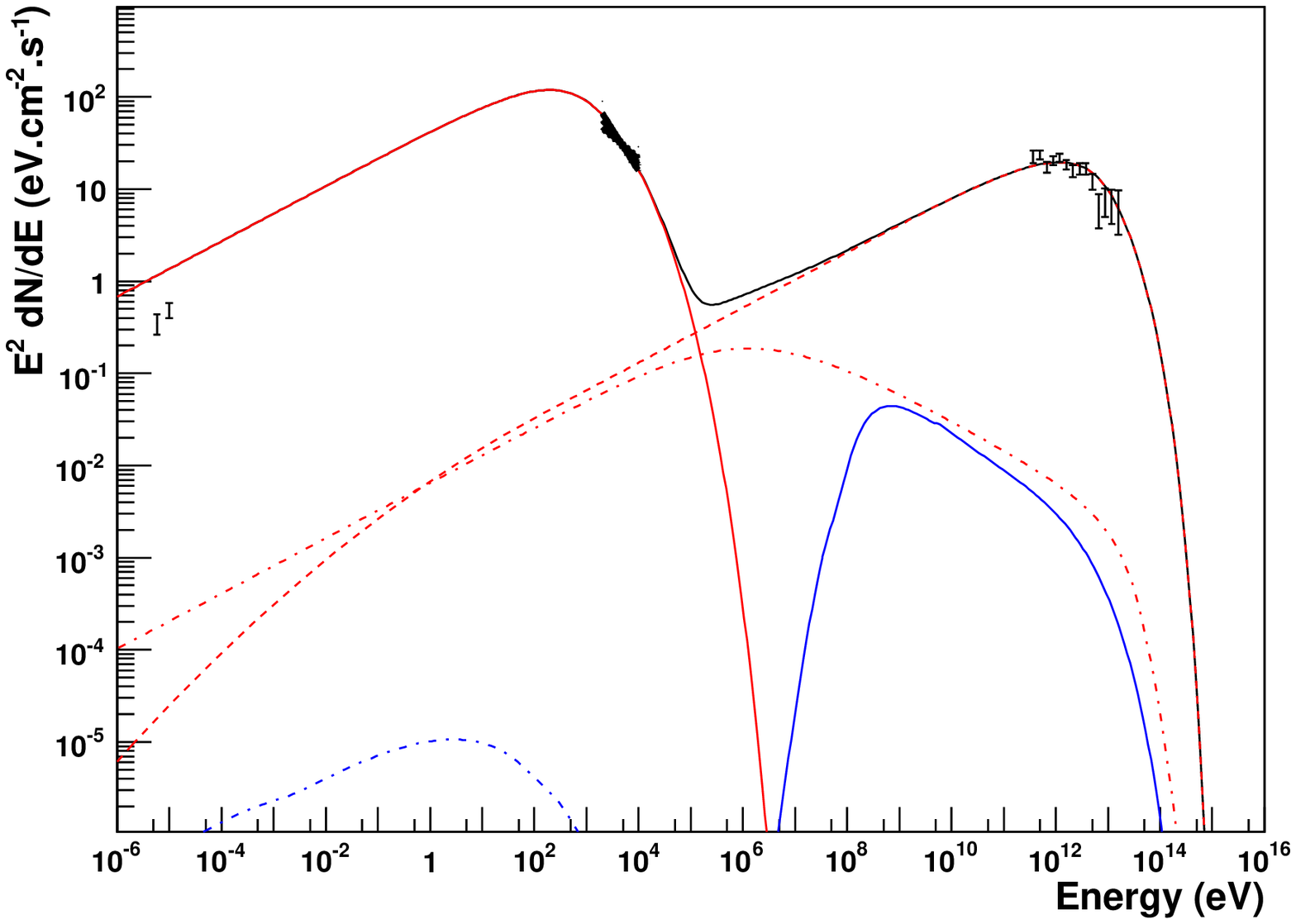}
\end{minipage}
\caption{Broadband SED models of RX~J0852.0-4622 for an electronic scenario in the nearby (left) and distant (right) case.
a) On the left, the modeling was done by using an injection spectrum in the form of a power-law of index 2.4, an exponential cut-off at 40~TeV and an electron/proton ratio $K_{ep} = 1.7 \times 10^{-3}$. The magnetic field amounts to $6 \, \mathrm{\mu G}$ and the density of the ambient medium is $0.008 \, \rm{cm^{-3}}$.
b) On the right, the modeling was done by using an injection spectrum in the form of a power-law of index 2.4, an exponential cut-off at 80~TeV and an electron/proton ratio $3.5 \times 10^{-2}$. The magnetic field amounts to $6.5 \, \mu {\rm G}$, and the density of the ambient medium is $0.01 \, \rm{cm^{-3}}$. 
The Parkes data~\citep{duncan} in the radio range, the ASCA data and the H.E.S.S. data are indicated. Red lines correspond to electrons, and blue lines to protons. The following processes have been taken into account: synchrotron radiation of primary (solid red line) and secondary (dotted blue line) electrons, IC scattering (dotted red line), bremsstrahlung (dotted-dashed red line) and proton-proton interaction (solid blue line).}
\label{fig:fitlepton}
\end{figure}

\begin{figure}[htbp]
\centering
\begin{minipage}{0.47\linewidth}
\centering
\includegraphics[width=\linewidth]{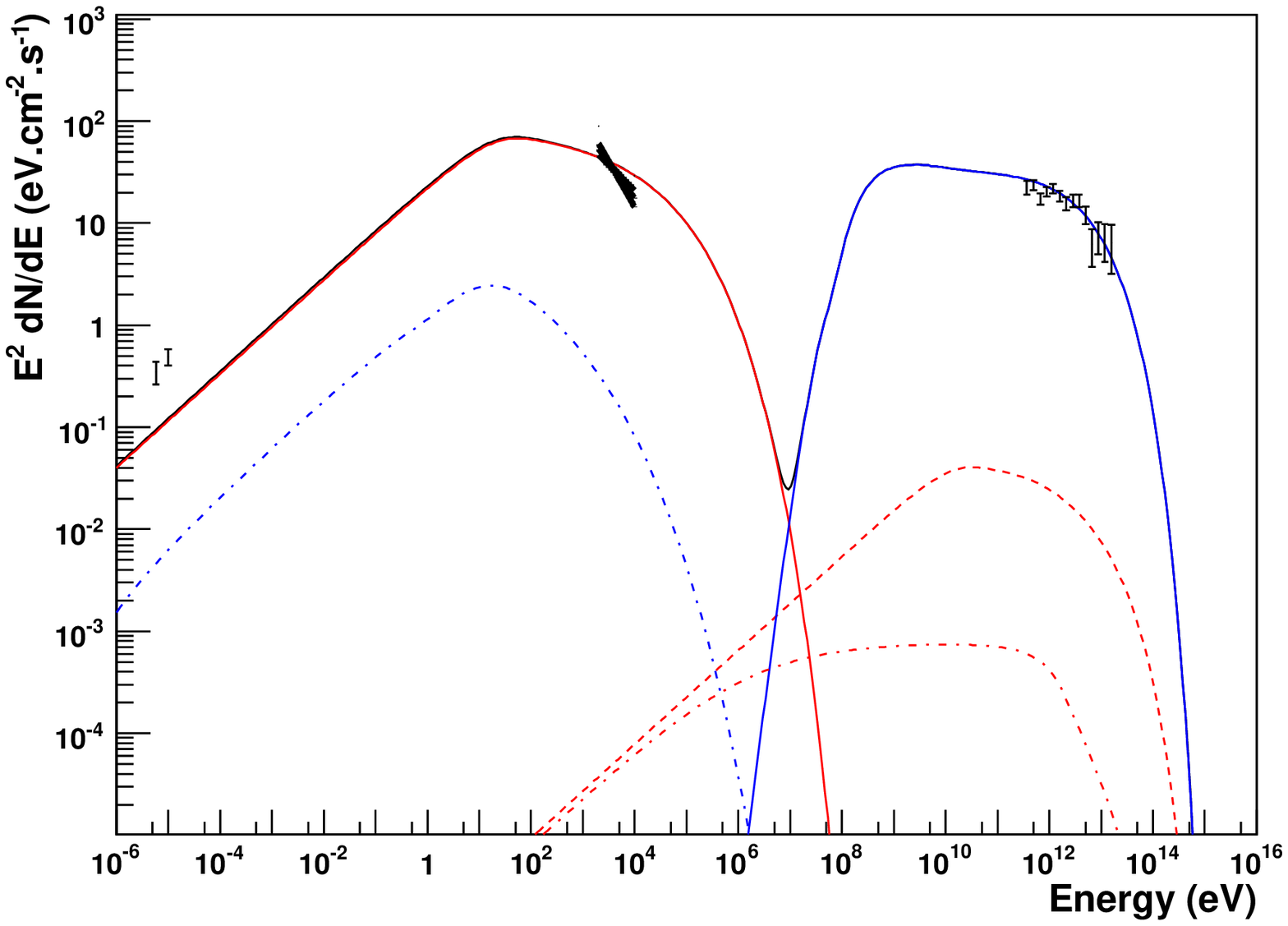}
\end{minipage}
\hfill
\begin{minipage}{0.47\linewidth}
\centering
\includegraphics[width=\linewidth]{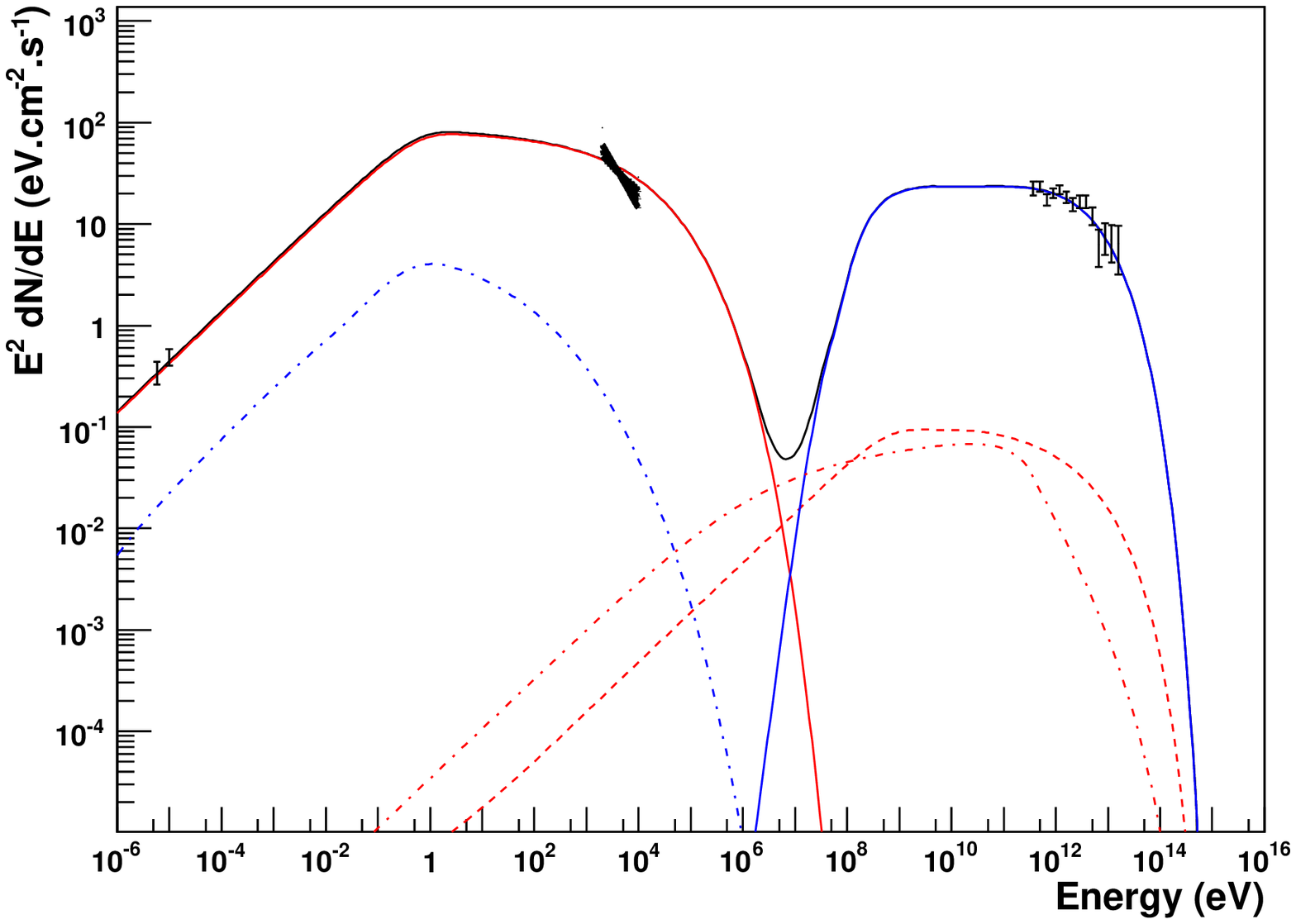}
\end{minipage}
\caption{Broadband SED models of RX~J0852.0-4622 for a hadronic scenario in the nearby (left) and distant (right) case.
a) On the left, the modeling was done by using an injection spectrum in the form of a power-law of index 2.1, an exponential cut-off at 110~TeV and an electron/proton ratio $2.4 \times 10^{-6}$. The magnetic field amounts to $120 \, \mu {\rm G}$, and the density of the ambient medium is $0.20 \, \rm{cm^{-3}}$.
b) On the right, the modeling was done by using an injection spectrum in the form of a power-law of index 2.0, an exponential cut-off at 100~TeV and an electron/proton ratio $4.5 \times 10^{-4}$. The magnetic field amounts to $85 \, \mu {\rm G}$, and the density of the ambient medium is $2.0 \, \rm{cm^{-3}}$.
The Parkes data~\citep{duncan} in the radio range, the ASCA data and the H.E.S.S. data are indicated. Red lines correspond to the electrons, and blue lines to protons. The following processes have been taken into account: synchrotron radiation of primary (solid red line) and secondary (dotted blue line) electrons, IC scattering (dotted red line), bremsstrahlung (dotted-dashed red line) and proton-proton interaction (solid blue line).}
\label{fig:fithadron}
\end{figure}

We have firmly established that the shell-type supernova remnant RX~J0852.0-4622 is a TeV emitter and for the first time we have resolved its morphology in the gamma-ray range. The thin shell observed by H.E.S.S. is highly correlated with the emission observed in X-rays with the ROSAT all-sky survey and ASCA but is also very similar to the morphology resolved in radio by ATCA. The overall gamma-ray energy spectrum extends over two orders of magnitude, providing the direct proof that particles of $\sim 100$~TeV are accelerated at the shock. There is an indication of deviation from a pure power law at high energy, but the lack of statistics does not enable us to draw any firm conclusions on this point. This spectrum is very similar to that of the other shell-type supernova remnant resolved by H.E.S.S., RX~J1713.7-3946, although the morphology of the latter was very different with a much thicker shell.\\
The question of the nature of the particles producing the gamma-ray signal observed by H.E.S.S. was also addressed. Despite the large uncertainty concerning the parameters of RX~J0852.0-4622, the H.E.S.S. data already give some strong constraints. In the case of a close by remnant, the results of the morphological study combined with our spectral modeling highly disfavour the electronic scenario which is unable to reproduce the thin shell observed by H.E.S.S. and the thin filaments resolved by Chandra. The hadronic scenario can approximately reproduce the data at the expense of a very low electron/proton ratio. In the case of a medium distance, the explosion energy needed to explain the gamma-ray flux observed by H.E.S.S., taking into account the limit on the density implied by the absence of thermal X-rays, would disfavour the hadronic process. At larger distances, both the electronic and the hadronic scenario are possible, at the expense, for the electronic process, of a low magnetic field of $\approx 6 \, \mu \rm G$. Such a small magnetic field exceeds typical interstellar values only slightly and is difficult to reconcile with the theory of magnetic field amplification at the region of the shock~\citep{bell}.\\
Finally, it appears clearly from Figures~\ref{fig:fitlepton} and \ref{fig:fithadron}, that the flux expected for lower energy gamma-rays ($E < 200$~GeV) for the electronic process (synchrotron + IC scattering) or for the hadronic process (proton-proton interactions) are significantly different. The results which should hopefully be obtained by INTEGRAL, GLAST or H.E.S.S. II will therefore have a great interest for the domain.

\acknowledgments
The support of the Namibian authorities and of the University of Namibia
in facilitating the construction and operation of H.E.S.S. is gratefully
acknowledged, as is the support by the German Ministry for Education and
Research (BMBF), the Max Planck Society, the French Ministry for Research,
the CNRS-IN2P3 and the Astroparticle Interdisciplinary Programme of the
CNRS, the U.K. Particle Physics and Astronomy Research Council (PPARC),
the IPNP of the Charles University, the South African Department of
Science and Technology and National Research Foundation, and by the
University of Namibia. We appreciate the excellent work of the technical
support staff in Berlin, Durham, Hamburg, Heidelberg, Palaiseau, Paris,
Saclay, and in Namibia in the construction and operation of the
equipment. M. Filipovic would like to thank 
Milorad Stupar for his work on the ATCA data.

\end{document}